\documentclass[prb,aps,twocolumn,floatfix,footinbib,superscriptaddress,showpacs,showkeys,notitlepage,amsmath,amssymb,amsfonts]{revtex4-1}
\usepackage{graphicx}
\usepackage{color}
\usepackage{ulem}
\usepackage{hyperref}
\usepackage{array}
\usepackage{multirow}
\begin{document}
\newcommand{\WW}{\mathbb{W}}
\newcommand{\CC}{\mathcal{C}}
\newcommand{\NN}{\mathcal{N}}
\newcommand{\dd}{$d$}
\newcommand{\pp}{$p$}
\newcommand{\sss}{$s$}
\newcommand{\ppp}{\boldsymbol{p}}
\newcommand{\ff}{$f$}
\newcommand{\UU}{$U$}
\newcommand{\WWW}{$W$}
\newcommand{\vv}{$v$}
\newcommand{\JJ}{$J$}
\newcommand{\uk}{$\mathcal{U}$}
\newcommand{\vk}{$\mathcal{V}$}
\newcommand{\jk}{$\mathcal{J}$}
\newcommand{\wk}{$\mathcal{W}$}
\newcommand{\imp}{\textrm{imp}}
\newcommand{\loc}{\textrm{loc}}
\newcommand{\zer}{\boldsymbol0}
\newcommand{\txr}[1]{\textcolor{black}{#1}}
\newcommand{\txrs}[1]{\textcolor{black}{\sout{#1}}}
\newcommand{\txb}[1]{\textcolor{black}{#1}}
\newcommand{\txbs}[1]{\textcolor{black}{\sout{#1}}}
\newcommand{\txg}[1]{\textcolor{black}{#1}}
\newcommand{\txm}[1]{\textcolor{magenta}{#1}}
\newcommand{\txms}[1]{\textcolor{magenta}{\sout{#1}}}
\newcommand{\txc}[1]{\textcolor{black}{#1}}
\newcommand{\txbr}[1]{\textcolor{black}{#1}}
\newcommand{\crpa}{cRPA}

\title{
Low-energy effective Hamiltonians for correlated
electron systems beyond density functional theory
}
\author{Motoaki Hirayama}
\affiliation{Department of Physics, Tokyo Institute of Technology, Ookayama, Meguro-ku, Tokyo 152-8551, Japan}
\author{Takashi Miyake}
\affiliation{Research Center for Computational Design of Advanced Functional Materials, AIST, Tsukuba 305-8568, Japan}
\author{Masatoshi Imada}
\affiliation{Department of Applied Physics, University of Tokyo, 
  Bunkyo-ku, Tokyo 113-8656, Japan}
\author{Silke Biermann}
\affiliation{
Centre de Physique Th\'eorique, 
Ecole Polytechnique, CNRS, Universit\'e Paris-Saclay,
F-91128 Palaiseau, France}

\date{\today}

\begin{abstract}
We propose a refined scheme of deriving an effective low-energy Hamiltonian
for materials with strong electronic Coulomb correlations
beyond density functional theory (DFT).
By tracing out the electronic states away from the target 
degrees of freedom in a controlled way by a perturbative scheme
we construct an effective model for a 
restricted low-energy target space incorporating the effects of
high-energy degrees of freedom in an effective manner.
The resulting effective model can afterwards be solved by accurate 
many-body solvers.
We improve this ``multi-scale {\it ab initio} scheme for correlated 
electrons'' (MACE) primarily in two directions:
(1) Double counting of electronic correlations between the DFT and the
low-energy solver is avoided by using the constrained GW scheme.
(2) The frequency dependence of the interaction emerging from the partial 
trace summation is taken into account as a renormalization to the 
low-energy dispersion.
The scheme is successfully tested on the example of SrVO$_3$.
Our work opens unexplored ways to understanding the electronic 
structure of strongly correlated systems beyond current DFT methods.
\end{abstract}


\maketitle

\section{Introduction}
\label{Introduction}
Strongly correlated electron systems are widely found in condensed matter
and have proven to generate many attractive phenomena and fundamental concepts
including quantum phase transitions and fluctuations 
such as superconducting and metal-insulator phenomena~\cite{imada98} with potential applications to future technology. 
Their accurate and {\it ab initio} theoretical treatment with predictive power
is therefore one of the grand challenges of nowadays condensed matter physics. 
However, conventional {\it ab initio} computational schemes based on 
density functional theory (DFT)~\cite{kohn} or 
many-body perturbation theory in the so-called
GW approximation~\cite{gw} are known to encounter
serious difficulties when electronic correlation effects are crucial.

Recently proposed versatile multi-scale methods~\cite{Review2010, biermann-review, kotliar-review}
that make use of the hierarchical energy structure of strongly correlated electrons 
(hereafter abbreviated as MACE, for
``multi-scale {\it ab initio} schemes for correlated electrons'') 
have opened the way to the design of increasingly accurate methods,
which are now overtaking the conventional ones. 
In strongly correlated electron systems in condensed matter,
the low-energy degrees of freedom (L part) 
represented by the bands near the Fermi level in the DFT or GW scheme are
typically sparse and isolated from 
the high-energy degrees of freedom (H part) obtained as dense bands 
away from the Fermi level.
This is not accidental because the isolation of the H and sparse L part is
a necessary condition for the strong electron correlation. (Otherwise,
it would be a weakly correlated system because of the screening by the H part or by self-screening by the L part. 
For a more extended discussion, see Ref.~\onlinecite{Review2010}).
MACE schemes take advantage of this hierarchical separation in the energy space
to treat the L part within highly accurate but relatively expensive numerical
techniques that could not be used directly for the full space in contrast to molecules and clusters treated in quantum chemistry~\cite{White, Kurashige,casula04} and nuclear physics~\cite{Barrett, Hjorth-Jensen}.
For the H part, on the other hand, cheaper techniques can be employed
thanks to less significant quantum fluctuations.

Motivated by the fact that the H part behaves effectively as an 
insulator with a gap around the Fermi level once the dynamics within the L part is excluded, one can treat
it in a controllable and accurate way by conventional methods such as the DFT within 
the local density approximation (LDA) or GW. 
This is not in contradiction to the fact that 
our target materials are strongly correlated electron 
systems and perturbative approaches do not work as a whole,
because strong correlation effects appear prominently in the excitations within the L part that will be treated and solved afterwards beyond the LDA or GW.   
Thanks to their mutual isolation, effects of the H part on the L part can be safely calculated perturbatively
in the spirit of the constrained random phase approximation 
(cRPA)~\cite{crpa,miyake08}:
A partial trace 
summation only over the H part is taken, effectively determining the
renormalization of the L part by the H part.

Physical properties of interest 
live in most cases on the energy scale of room temperature or below, and 
certainly within the L part. Therefore, an accurate treatment of the degrees 
of freedom in the L part is required.
This low-energy system (L part) can indeed be solved
using nonperturbative many-body tools such as quantum Monte Carlo (QMC) 
methods and various renormalization group schemes~\cite{White,Kashima,Hallberg,Kurashige}
-- in particular the variational Monte Carlo (VMC)~\cite{Gros,Tahara,casula04,Review2010} 
-- or dynamical mean field theory (DMFT), and related methods~\cite{Georges96,kotliar-review,DCA,biermann-review,held-review} obtaining
results beyond the DFT or standard GW schemes by the combination of the DFT and DMFT (DFT+DMFT)\cite{Anisimov97, Lichtenstein98}.

Such an {\it ab initio} hierarchical scheme
has proven useful and successful for a wide range of materials questions~\cite{Review2010, Review_SB2014},
from transition metals~\cite{Mn,Fe,solovyev,Braun,Sanchez}, 
their oxides~\cite{imai,imai2,yvo3,biermann_vo2,v2o3,aryasetiawan09,Thunstrom}, sulphides~\cite{bavs3}, pnictides~\cite{nakamura08,aichhorn09,miyake10,aichhorn10,misawa11JPSJ,misawa12PRL,haule11,Platt_H_T,misawa14NComm},
rare earths~\cite{ce, bieder,sorella15} and
their compounds~\cite{ce2o3, cefeaso, cefeaso2, cesf},
including heavy fermions~\cite{CeIrIn5, CeCu2Si2},
actinides~\cite{Pu, Pu-Savrasov, Am} and their compounds~\cite{Kolorenc}
to organics \cite{nakamuraET,organics,nakamura12,shinaokaET}, correlated semiconductors \cite{gaas,FeSb}, spin-orbit materials \cite{sriro, sriro-arita,yamaji14} and correlated surfaces and interfaces~\cite{lao-sto,hansmann}.
In this hierarchical scheme such as the DFT+DMFT, an effective Hamiltonian within a low-energy
window around the Fermi level is obtained using the DFT, and this
Hamiltonian is then solved by a low-energy solver.
This construction thus makes explicit use of the ``separability''
of high- and low-energy degrees of freedom.
However, in most current schemes, little effort is devoted to
the electronic structure of the higher energy degrees of freedom
--  which are simply described at the DFT level --
and their influence on the low-energy part.

In practice, examples where electron correlation effects were overestimated have also been found: a typical case are organic conductors~\cite{nakamuraET,shinaokaET}, 
where many-body calculations using cRPA values for the interaction -- even after ``dimensional downfolding''~\cite{NakamuraD} -- overestimate correlation effects as compared to experiments.
Along the same lines, it has been argued that, in iron
pnictide compounds, the ratio of the effective Coulomb interactions as estimated
within the cRPA to the effective bandwidth of the Kohn-Sham band structure
of the DFT is slightly overestimated as compared to experimental results~\cite{misawa11JPSJ,misawa12PRL,misawa14NComm}. 
On the other hand, it is known that the neglect of dynamical effects in the screening by the H part leads to an underestimation of correlation effects~\cite{werner12} (since the effective bandwidth is overestimated)~\cite{casula-ZB}. It was proposed that this subtlety has relevance to the low-temperature metallic and nonmagnetic state of FeSe as well as the so-called bicollinear antiferromagnetic order of FeTe~\cite{Hirayama2015}. 
A further example is the transition metal pnictide
BaCo$_2$As$_2$ where dynamical screening effects have been invoked
to explain the puzzling absence of ferromagnetism despite a large LDA density
of states at the Fermi level~\cite{vanroeke2014}.
We will come back to these observations in the discussion section
at the end of this paper.

In MACE schemes, the accurate derivation of the effective low-energy 
models is crucially important for the quantitative level of the predictive 
power of the calculations. Therefore, the DFT or GW calculations for the 
global electronic structure including both the L and H part must be 
consistently bridged to the effective models in the L part. 
The main challenge consists in avoiding double counting of the electronic
correlations and screening 
already taken into account at the DFT or
GW level: In the DFT, Coulomb interactions are treated through the construction
of an effective potential, the (Kohn-Sham) exchange correlation potential,
while the GW scheme constructs a frequency-dependent many-body self-energy
(albeit in a perturbative manner). On the other hand, the low-energy effective models 
are solved by low-energy solvers, where the electron correlation effects are 
more accurately treated within this low-energy degrees of freedom.
Therefore, there exists overlap in treating the low-energy part of the electron correlation.
This is  known as the ``double counting problem", and
a careful and improved treatment to 
avoid double counting is required.

At the DFT level, the nonlinear dependence of the exchange-correlation
potential on the electronic density makes the formal separation of 
the correlation energy contributions stemming from a subset of orbitals 
an ill-defined problem. 
On a conceptual level, strictly double counting-free schemes are therefore
only possible when avoiding the use of the DFT altogether.
Double-counting-free schemes can be defined e.g. based on many-body
perturbation theory: in this case, the exchange-correlation potential
is replaced by a perturbative self-energy, calculated directly in a
Green's function language. The combined ``GW+DMFT'' scheme~\cite{gwdmft}
illustrates the advantages of such an approach. 
A simpler scheme, derived from the GW+DMFT, is the recently proposed 
``screened exchange dynamical mean field theory'' ~\cite{vanroeke2014, epl2014}, where
the DFT exchange-correlation potential is eliminated and
replaced by a screened exchange term.
We also mention a recent attempt to transfer this concept to 
the LDA+DMFT scheme~\cite{haule}.

These considerations motivate the construction of effective low-energy
Hamiltonians based on many-body perturbation theory for the H part, rather than on
the Kohn-Sham Hamiltonian of the DFT. As we already mentioned, the perturbative treatment of the H part is indeed justified by the fact that the exclusion of excitations within the L part makes the system ``insulating" with the suppressed vertex correction.

A disadvantage is, however,
that the resulting effective models are naturally first given as
models with frequency-dependent parameters.
Indeed, integrating out the high
energy degrees of freedom generically generates frequency-dependent
effective interactions and hopping. It is very useful to further 
reduce such models to effective Hamiltonian forms because 
low-energy solvers for frequency-independent Hamiltonians are computationally
less demanding.
In this paper, we develop a consistent and accurate {\it ab initio} 
framework of deriving the low-energy effective Hamiltonian of the L part
in view of the construction of a complete MACE scheme.
Our aim in the present article is to derive 
low-energy effective models for the L-space
that are as accurate as possible and can be
treated by sophisticated low-energy solvers in the subsequent step.

The paper is organized as follows: In Section~\ref{general}, we
describe the general strategy to be employed. In section~\ref{Outline}, we give a first outline of the equations,
followed by the detailed derivations in Section~\ref{derivation}.
Section~\ref{summary} presents a concise summary of the obtained
scheme, while Section~\ref{variants} explains variants of the
scheme. In Section~\ref{results}, we present calculations on
the perovskite oxide SrVO$_3$, illustrating how the scheme works
and giving practical information on the relative importance of the
different terms. Finally, we present our conclusions and perspectives
in Section~\ref{conclusion}.

\section{Outline of derivation of low-energy effective Hamiltonian}
\label{Outline}

\subsection{General strategy}
\label{general}
We decompose the full Hilbert space into low-energy (the L space)
and high-energy subspaces (the H part) thanks to the hierarchical
structure of strongly correlated electron systems as 
described  in Sec.~\ref{Introduction}.
Our construction for the decomposition and bridging 
between the two parts
will be based on the GW scheme, since this allows
for a well-defined way to avoid double-counting. Indeed, at the
DFT level, after identifying the H- and L-spaces,
the partial trace summation and the elimination of the H-space 
can be performed by means of the cRPA~\cite{crpa,Review2010}, as 
described below. However, when the low-energy part is 
solved by a refined many-body solver, some part of the interactions are
counted twice since the initial DFT calculation
already contains the correlation effects for the L part.
Indeed, the DFT considers the exchange correlation contribution 
without distinguishing the L- and H-spaces and it is impossible
to disentangle the two spaces at this level.
The GW scheme, instead, allows for the subtraction of the double counting
by calculating a constrained self-energy as constructed in
Ref.~\onlinecite{Hirayama}.
This constrained self-energy incorporates the interactions in the
form of a self-energy 
from which the contribution of the L part 
has been excluded.

After eliminating 
the H-space, the L-space is expressed 
by single-, two-, three-particles and even higher terms.  However, the effective many-body 
interaction higher than the two-particle channel is expected to be small 
if the target L-space is isolated from the H-space. This is 
true in typical strongly correlated systems, and motivates a perturbative
treatment of the H-L-coupling. 
In this article, 
we ignore multi-particle effective interactions of higher order than the two-body terms. 

The single-particle (kinetic energy) terms are modified (renormalized) 
by the constrained self energy. 
The two-particle (effective Coulomb interaction) terms are represented 
as the partially screened interaction obtained from the cRPA~\cite{crpa}. 
In general, the self-energy and the screened interaction 
are frequency-dependent, thus not allowing for a representation 
in a Hamiltonian form.

As mentioned above, in this paper, we focus on methods that derive
the low-energy effective models described in the form of Hamiltonians 
\begin{eqnarray}
H_{\rm eff}=\sum_q T_{\rm eff}(q)c^{\dagger}_qc_q+\sum_{q,k,p} W_{\rm H}(q)c^{\dagger}_k c_{k+q}c^{\dagger}_p c_{p-q},
\label{zeroth-Heff}
\end{eqnarray}
where 
the renormalized single particle dispersion $T_{\rm eff}(q)$ after incorporating the self-energy effect and the effective
interaction $W_{\rm H}(q)$ screened by the H part constitute the model for the 
electrons in the L part represented by the creation (annihilation) operators for the
electron, $c_k^{\dagger} (c_k)$ at momentum $k$. Here, for simplicity,
spin and orbital indices are omitted.

Our task at this stage is thus to map the model with frequency dependent
single- and two-particle terms onto a frequency-independent Hamiltonian 
in a controlled way. For this mapping, we propose a scheme that combines 
the merits of the works by Hirayama {\it et al.}~\cite{Hirayama} and 
Casula {\it et al.}~\cite{casula-ZB}: In a step-by-step procedure,
we include the influence of the H-space into the L-space and eliminate
the frequency dependence by taking into account its effect on the
Hamiltonian in the form of an effective renormalization of the parameters.

We remark that, in practice, 
the derived effective low-energy Hamiltonian satisfies the following principle:
If one solved the effective low-energy Hamiltonian within the GW-type perturbative 
treatment instead of the accurate low-energy solver, that would yield the same result 
as the solution obtained by the same perturbative scheme starting from the full space including the H and L parts. 
This is called the ``chain rule", which justifies the effective Hamiltonian as that for the L part.

Renormalized single-particle Hamiltonian
Our starting point for the single-particle part is the DFT band dispersion
denoted by $\epsilon_{\rm DFT}(q)$ and the corresponding 
Kohn-Sham Hamiltonian $H^{(0)}=\sum_q \epsilon_{\rm DFT}(q)c^{\dagger}_qc_q$.
Here, $c^{\dagger}_q (c_q)$, is the creation (annihilation) operator of an 
electron with wavevector $q$. We have suppressed the spin and band 
indices for simplicity.
Then, the single particle Green's function 
$G_{\rm DFT}$ reads
\begin{eqnarray}
G_{\rm DFT}(q,\omega)=1/[\omega-\epsilon_{\rm DFT}].
\end{eqnarray}

On the DFT level, electronic correlations are taken into
account in the form of an effective exchange correlation potential 
$V_{\rm xc}(q)$. As discussed above, treating the electron correlation 
effects in the L-space explicitly within the low-energy solver would
lead to a double counting of electronic correlation in the L-space.
To avoid the double counting, $V_{\rm xc}(q)$ is subtracted and replaced
by a corrective self-energy $\Delta \Sigma (q,\omega)$.
By incorporating $\Delta \Sigma$, the effective single-particle part reads 
\begin{eqnarray}
T_{\rm eff}(q,\omega)=\epsilon_{\rm DFT}-V_{\rm xc}(q)+\Delta \Sigma (q,\omega).
\label{Teff(q,w)}
\end{eqnarray}

\subsection{H-space contribution to self-energy: constrained self-energy}
$\Delta \Sigma$ comes from two contributions; $\Delta\Sigma=\Delta\Sigma_{\rm H}+\Delta\Sigma_{\rm L}$. 
$\Delta\Sigma_{\rm H}$ is the contribution to the self-energy from the H space, while
$\Delta\Sigma_{\rm L}$ is from the frequency dependent part of the effective interaction
incorporated into the self-energy, which is the constrained self-energy effect within the L space obtained by excluding the self-energy arising from the static effective interaction $W_{\rm H}(q)$. 
Here, we sketch the idea for $\Delta\Sigma_{\rm H}$ and discuss $\Delta\Sigma_{\rm L}$ in the next subsection.
A specific form for the correction $\Delta\Sigma_{\rm H}$, dubbed ``constrained self-energy'',
was already derived in Ref.~\onlinecite{Hirayama}, based on a restricted GW calculation.
The basic prescription is to add only the self-energy arising from the contribution 
of the H-space, by excluding the part stemming purely from the L-space. 
The reason why one should exclude the self-energy stemming from the 
L-space is that this part is more accurately calculated within the 
low-energy solver afterwards.

\subsection{Renormalization to self-energy from frequency-dependent partially-screened interaction }
The cRPA~\cite{crpa} was 
proposed as a means to calculate the effective local Coulomb interactions
to be used in the L-space from a systematic first-principles procedure.
It can be understood as a way of tracing out the H-space for deriving the effective interaction, while
keeping track of the resulting renormalization of the L-space
degrees of freedom.
The tracing out of the H-space by the standard cRPA results in an effective interaction
for the two-particle part of the model in the L-space in the form
\begin{eqnarray}
W_{\rm H} (q,\omega) = \frac{v(q)}{1- P_{\rm H}(q,\omega)v(q)},
\label{Wr_crpa}
\end{eqnarray}
where the wave-number ($q$) dependent bare Coulomb interaction $v$ 
is partially screened by the partial polarization $P_{\rm H}$. 
Here, $P_{\rm H}$ is defined in terms of the total polarization $P$ by excluding 
the intra-L-space polarization $P_{\rm L}$: $P_{\rm H}\equiv P-P_{\rm L}$.
$P_{\rm L}$ involves only screening processes within the L-space.

\begin{figure}[h]
\centering 
\includegraphics[clip,width=0.45\textwidth ]{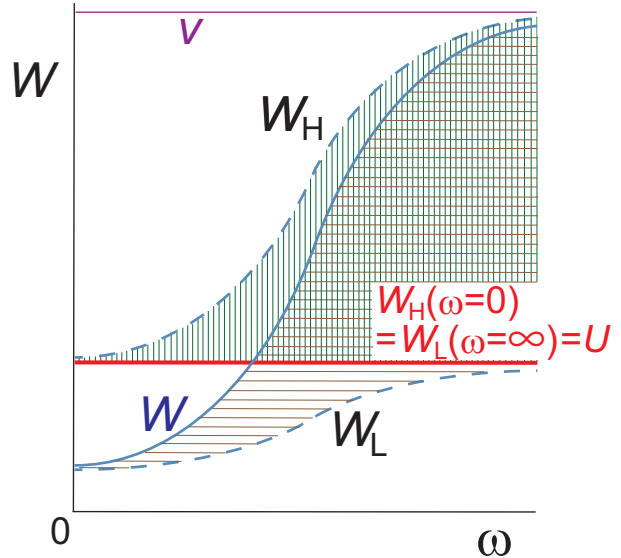} 
\caption{(Color online)
Schematic frequency dependence of effective interaction screened
from bare interaction $v$, and  
obtained from full RPA (GW) ($W$), cRPA ($W_{\rm H}$) and screened interaction by RPA ($W_{\rm L}$) within low-energy effective model at the effective interaction $U=W_{\rm H}(\omega=0)$.
This is only a qualitative feature and more realistic dependence is seen in Fig.~\ref{W}.
}
\label{Cartoon_W}
\end{figure}

Here, $W_{\rm H}$ is frequency dependent as
schematically illustrated in Fig.~\ref{Cartoon_W}.
However, most many-body calculations in the literature that use the effective 
interactions from the cRPA method or similar schemes neglect this
frequency-dependence (exceptions are 
Refs.~\onlinecite{werner12},
\onlinecite{casula-ZB},
\onlinecite{vanroeke2014}, 
\onlinecite{Hirayama},
\onlinecite{casula-prb},
\onlinecite{werner-LCO}, 
\onlinecite{cafe2as2},
\onlinecite{sakuma2013}, 
\onlinecite{sakuma14},
\onlinecite{tomczak-gwdmft},
\onlinecite{TomczakPRB2014}),
and use only the zero-frequency value of the
interaction $W_{\rm H}(q,\omega=0)$ for the construction of the low-energy effective Hamiltonian~\cite{crpa,Hirayama}.

We note that this static limit $W_{\rm H} (q,\omega=0) $ obtained by using Eq.~(\ref{Wr_crpa}) in fact
satisfies the above mentioned chain rule: 
The whole dynamical interaction emerging when 
one solves the whole H- and L-space degrees of freedom by RPA is the
usual fully screened interaction 
$W(q,\omega)$ given by
\begin{eqnarray}
W (q,\omega) = \frac{v(q)}{1- P(q,\omega)v(q)},
\label{full_rpa}
\end{eqnarray}
as is depicted in Fig.~\ref{Cartoon_W}.

On the other hand, 
if we calculate the screening by the RPA within the L-space by regarding as if $W_{\rm H} (q,\omega=0)$ 
would be the bare interaction, 
this leads to a screened interaction
\begin{eqnarray}
W_{\rm L} (q,\omega) = \frac{W_{\rm H}(q,\omega=0)}{1- P_{\rm L}(q,\omega)W_{\rm H}(q,\omega=0)},
\label{W_L}
\end{eqnarray}
which is depicted schematically in Fig.~\ref{Cartoon_W}.
Here, $P_{\rm L}$ is the RPA polarization in the low-energy
subspace.
Note that $W_{\rm L} (q,\omega\rightarrow \infty) = W_{\rm H}(q,\omega=0)$.
Then the chain rule
$W_{\rm L} (q,\omega) =W(q,\omega=0)$ can be proven~\cite{crpa}.

However, the static $W_{\rm H}(q,\omega=0)$ amounts to neglecting the frequency dependent part 
\begin{eqnarray}
W^{\rm dyn}_{\rm H}(q,\omega)\equiv W_{\rm H}(q,\omega)- W_{\rm H}(q,\omega=0)
\label{Wrdyn}
\end{eqnarray}
depicted by the vertical hatching in Fig.~\ref{Cartoon_W}.
In this paper, we will take into account the 
contribution of this dynamical part as the renormalization to the 
kinetic energy part, either as a perturbative
self-energy or in a nonperturbative fashion.
In the effective low-energy model, we then keep $W_{\rm H}(q,\omega=0)$ for the effective interaction. 
In the case of the perturbative treatment, for example, 
the contribution to the self-energy $\Delta \Sigma_{\rm L}$ is 
\begin{eqnarray}
\Delta \Sigma_{\rm L}^{\rm Pert}=G_{\rm L} W^{\rm dyn}_{\rm H}(q,\omega)
\label{DeltaSigmaH}
\end{eqnarray}
as was formulated in Ref.~\onlinecite{Hirayama} and we review in detail in the next section.

We also remark that the dynamical part to be considered can be improved
from Eq.~(\ref{Wrdyn}) in a more consistent manner:
Since the screening on the RPA level within the L-space is $W_L$ in Eq.~(\ref{W_L}), 
one realizes that the dynamical part of the interaction
ignored when we use the low-energy solver is   
\begin{eqnarray}
W^{\rm dyn}_{\rm GW}(q,\omega)\equiv W(q,\omega) - W_{\rm L} (q,\omega)
\label{WdynGW}
\end{eqnarray}
as depicted as the horizontal hatching in Fig.~\ref{Cartoon_W}. 

Then we need to take into account the renormalization (namely, self-energy effect) originating from $W^{\rm dyn}_{\rm GW}(q,\omega)$ instead of $W^{\rm dyn}_{\rm H}(q,\omega)$.
The perturbative contribution to the self-energy then replaces Eq.~(\ref{DeltaSigmaH}) with  
\begin{eqnarray}
\Delta \Sigma_{\rm L}^{\rm GW}=G_{\rm L} W^{\rm dyn}_{\rm GW}(q,\omega).
\label{DeltaSigmaGW}
\end{eqnarray}
Using $W^{\rm dyn}_{\rm GW}(q,\omega)$ in Eq.~(\ref{DeltaSigmaGW}) replacing $W^{\rm dyn}_{\rm H}(q,\omega)$ is expected to improve the self-energy, because it takes into account the missing part of the
L-space dynamics on the GW level and satisfies the chain rule 
even for the self-energy as we show in the following.

A conventional one-shot GW calculation in the full space gives
full self-energy from the fully screened Coulomb interaction $W$ (Eq.~(\ref{full_rpa})) as
\begin{eqnarray}
 \Sigma = G_{\rm L} W
 \label{full-self-energy}
\end{eqnarray}
(Note: we write here and in the following
symbolically $GW$. Depending
on if the calculation is done on the real/Matsubara
axis, a factor -1 or i has to be added.)
On the other hand, the self-energy within the L-space at the one-shot GW level
is 
\begin{eqnarray}
 \Sigma_{\rm L} = G_{\rm L} W_{\rm L}.
 \label{L-self-energy}
\end{eqnarray}
Then $\Sigma= \Delta \Sigma_{\rm L}^{\rm GW}+\Sigma_{\rm L}$
is obviously satisfied, which is the chain rule for the self-energy.
In this article, we compare the results calculated from the two choices, Eq.~(\ref{Wrdyn}) and Eq.~(\ref{WdynGW}) in examples to gain into physical insights.

The effective model is then given by 
\begin{eqnarray} 
H_{\rm eff}&=&\sum_q T_{\rm eff}(q,\omega)c^{\dagger}_qc_q \nonumber \\
&+&\sum_{q,k,p} W_{\rm H}(q,0)c^{\dagger}_k c_{k+q}c^{\dagger}_p c_{p-q},
\label{Heff_with_w}
\end{eqnarray}

with $T_{\rm eff}$ given by Eq.~(\ref{Teff(q,w)}) by employing either $\Delta \Sigma_{\rm L}^{\rm Pert}$ or $\Delta \Sigma_{\rm L}^{\rm GW}$ for $\Delta \Sigma_{\rm L}$ contained in $\Delta \Sigma$ in Eq.~(\ref{Teff(q,w)}).
Equation~(\ref{Heff_with_w}) still contains the frequency dependence in $T_{\rm eff}$,
which should be incorporated in the frequency-independent form by including the renormalization effect.

Then, on top of the zero-frequency limit $T_{\rm eff}(q,\omega=0)$, 
to incorporate the effects of the frequency-dependence,  
we implement the following procedure: 
First, the frequency dependence in the nonlocal part of $W_{\rm H}$ 
is taken into account perturbatively. 
This is done by constructing a self-energy
$\Delta \Sigma (q,\omega)$ along the lines of 
Ref.~\onlinecite{Hirayama}. This 
proposal employs $\Delta \Sigma_{\rm L}(q,\omega)=G_{\rm L}W^{\rm dyn}$,
where $W^{\rm dyn}$ is either $W^{\rm dyn}_{\rm H}$ or  $W^{\rm dyn}_{\rm GW}$. Thus incorporated renormalized 
single-particle part is linearized in 
$\omega$ as $T_{\rm eff}(q,\omega=0)-[d\Delta \Sigma (q,\omega)/d\omega]\omega$,
and the $\omega$ dependence is absorbed into the renormalization factor  
$Z_{\rm corr}=1/[1-d\Delta \Sigma (q,\omega)/d\omega|_{\omega=0}]$,
where the dispersion $T_{\rm eff}^{(0)}(q)$ is replaced with $Z_{\rm corr}T_{\rm eff}^{(0)}(q)$. 

The effects of the local part of the interaction are taken into 
account following the proposal by Casula {\it et al.}~\cite{casula-ZB}.
There it was shown that -- in the antiadiabatic limit --
a model with frequency-dependent interactions
can be mapped onto a model with static interactions and a renormalized
one-body Hamiltonian. The renormalization factor $Z_B$ can be explicitly
obtained from the frequency-dependence of the interaction. We will 
give the explicit form in the next section.

As a result, the single-particle part of the effective Hamiltonian 
Eq.~(\ref{zeroth-Heff}) as defined by its hopping $T_{\rm eff}^{(0)}(q)$ 
is replaced by a single-particle Hamiltonian with effective hopping 
\begin{eqnarray}
T_{\rm eff}^{(1)}(q)=[\epsilon_{\rm DFT}-V_{\rm xc} +
\Delta \Sigma (q,\omega=0)]Z_{\rm corr}Z_B,
\end{eqnarray}
which replaces $T_{\rm eff}(q,\omega)$ in Eq.~(\ref{Heff_with_w}) and the Hamiltonian form (\ref{zeroth-Heff}) is obtained.

\section{Detailed Derivation of Effective Models}
\label{derivation}

In this section, we give a detailed description of the derivation 
of the effective models outlined above.~\subsection{Starting point}

We start from the ``non-interacting''
Hamiltonian $H^{(0)}(k)$,
and assume that we are
working in a basis where its single-particle part 
is block diagonal at each $k$-point (e.g. 
in a Wannier gauge associated with atom-centred
Wannier functions constructed separately for the
L- and H-spaces). 
We will think of $H^{(0)}$ as the Kohn-Sham Hamiltonian of the DFT,
even though other choices are possible.
We assume that the block diagonality should be a good 
starting point, because
in many typical correlated materials such as typical 
transition metal oxides, the bands that have dominantly 
the character of the localized orbitals are energetically 
separated from itinerant bands such as ligand bands.
This fact helps the construction of effective models
since it implicitly guarantees the existence of such a
basis set. Indeed, vertex corrections that would mix
the two spaces decrease with the energetic separation.

A consequence of the block diagonality is that the
non-interacting Green's function $G^{(0)}$ is also
block-diagonal and can be decomposed into
\begin{eqnarray}
G^{(0)} = G^{(0)}_{ll} |L \rangle \langle L |
+ G^{(0)}_{hh} |H \rangle \langle H |
\end{eqnarray}
where the bra-kets are a shorthand for projectors
onto the respective subspaces.

We stress that
$\Sigma$ in Eq.~(\ref{full-self-energy}) is not in general block-diagonal.
Rather, it has both off-diagonal and diagonal
components, e.g.:
\begin{eqnarray}
 \Sigma_{lh} = G^{(0)}_{ll} W_{lllh} + G^{(0)}_{hh} W_{lhhh} -V_{xc lh}
\label{Sdr}
\\  
\Sigma_{ll} =  G^{(0)}_{ll} W_{llll} + G^{(0)}_{hh} W_{lhhl} -V_{xc ll}
\label{Sdd}
\end{eqnarray}
Here, $G^{(0)}_{ab} = - \langle T c_{a}(\tau) c^{\dagger}_{b}(0) \rangle$,
where $a, b$ denote elements of the H- or L-spaces, and 
$W_{abcd}$ 
is the coefficient of the interaction term $c^{\dagger}_ac_bc^{\dagger}_cc_d$. 
In the following, we will use the convention that $l$ and $h$ represent 
degrees of freedom belonging to $L$ and $H$ degrees of freedom, respectively. 

This matrix is used to calculate the interacting Green's
function
\begin{eqnarray} 
G(q,\omega)&=&\left[
G^{(0) -1} - \Sigma\right]^{-1}.
\label{G1}
\end{eqnarray}
Straightforward matrix inversion gives for the ll-block (low-energy block)
of this Green's function:
\begin{eqnarray} 
G(q,\omega)_{ll}&=&\frac{1}{
G_{ll}^{(0) -1} -  \Sigma_{ll}
-  \Sigma_{lh} G_{hh}  \Sigma_{hl}}.
\label{Gd}
\end{eqnarray}
In the following, we will use this form to extract a corrective
self-energy: The latter is given by 
those parts of $\Sigma_{ll} + \Sigma_{lh} G_{hh}  \Sigma_{hl}$
that are generated by the presence of the H-space.
This self-energy contribution should be taken into account at the
level of the construction of the low-energy effective model, as
an effective renormalization of the L-space by the H-space.

\subsection{Interspace exchange term}

The last term in the denominator of (\ref{Gd})
is an ``interspace exchange'' self-energy contribution
originating from the block-off-diagonal self-energy
$ \Sigma_{lh}$ ($ \Sigma_{hl}$) between L- and H-electrons. 
While it can in principle be treated by a direct calculation, we
prefer to disregard it at this stage. The reason is
that, within the low-energy subspace, it is in fact
a higher (second) order contribution in the 
interspace interaction. The interspace exchange interaction
is expected to be small if the H- and L-spaces are well separated.
In addition, the interspace exchange may at least partially 
cancel with the first order vertex term.

\subsection{Direct H-space contribution to constrained self-energy}

The corrective self-energy 
that we are interested in
here is thus contained in 
the second but last term in the denominator
of (\ref{Gd}), the block-diagonal self-energy 
$\Sigma_{ll}$ given by (\ref{Sdd}).
This quantity includes some
influence of the high-energy H-space through (a) the
screened Coulomb interaction $W_{llll}$ in the first term and (b) the entire second term. Here, $W_{llll}$ is either $W_{\rm H}(q,\omega)$ in Eq.~(\ref{Wr_crpa}) or $W(q,\omega)$ in Eq.~(\ref{full_rpa})
and the first term contains $\Delta\Sigma_{\rm L}$. 
This former part will be discussed in the next subsection. 
The latter gives
\begin{eqnarray}
\Delta \Sigma_{\rm H}(q,\omega) = G^{(0)}_{hh} W_{lhhl}.
\end{eqnarray}
As we will see below its effect is a band narrowing with respect
to the Hartree band structure, comparable to the effect of the
exchange-correlation potential $V_{xc}$ of the DFT.

This correction can either be applied directly as a frequency-dependent
additional self-energy term $\Delta \Sigma_{\rm H}(q,\omega)$, 
in which case it leads to a dynamical low-energy model, or one can
use a Taylor expanded approximate form.
If the low-energy behavior 
is to a good approximation linear, that is, its 
frequency dependence is well approximated as 
\begin{eqnarray}
\Delta \Sigma_{\rm H}(q,\omega) = \Delta \Sigma_{\rm H}(q,\omega=0)+ \Delta \Sigma_{\rm H}^{'}|_{\omega=0}\omega,
\end{eqnarray}
where $\Delta \Sigma_{\rm H}^{'}=d\Delta \Sigma_{\rm H}/d\omega$,
then the renormalization factor resulting from this contribution
is given by
\begin{eqnarray}
Z_{\rm H} = \frac{1}{1- \frac{\partial \Delta \Sigma_{\rm H} (q,\omega)}
{\partial \omega}|_{\omega=0}}.
\label{ZH}
\end{eqnarray}
At this level the effective kinetic energy is renormalized to 
\begin{eqnarray}
T_{\rm eff}^{(1)}(q)=[\epsilon_{\rm DFT}-V_{\rm xc} +\Delta \Sigma_{\rm H} (q,\omega=0 ) ]Z_{\rm H}.
\label{T:Sigma_H}
\end{eqnarray}

\subsection{Frequency-dependence of interactions within low-energy
space: non-local part}

We finally analyze the remaining term $G^{(0)}_{ll} W_{llll}$, the 
first term in Eq.~(\ref{Sdd}). The low-energy effective model has
to be constructed in such a way that -- at the GW level within
the L-space -- this self-energy would be reproduced.
The influence of the H-space, contained in this term through the
matrix element of the fully screened interaction $W$ or $W_{\rm H}$ hereby has
to enter in an effective way.

This can be naturally achieved when constructing a model with
the dynamical interaction $W_{\rm H}(q, \omega)$ as given in Eq.~(\ref{Wr_crpa}).
One thus obtains at first a model with nonlocal and frequency-dependent
interactions, and the
task is to map this model onto a frequency-independent 
Hamiltonian form by effectively renormalizing the 
Hamiltonian parameters. For that purpose, we treat the nonlocal and local 
parts of the interaction (two-body) terms separately.
We first eliminate 
the frequency dependence in the nonlocal part
by treating it within the perturbative scheme proposed by 
Hirayama {\it et al.}~\cite{Hirayama}.  The perturbative
treatment is justified, because the corresponding correction is small.
On the other hand, the local and frequency-dependent part can be large 
and we will treat it nonperturbatively
in the formalism proposed by Casula {\it et al.} \cite{casula-ZB}. 
This procedure allows for a nonperturbative treatment but
is only suitable for local interactions.
An additional subtlety arises due to the fact that the nonperturbative
treatment does not take on the form of a self-energy but rather a
direct renormalization of the hopping. Therefore, no zero-frequency
part appears in the procedure by Casula {\it et al.}, and we therefore
retain the local static part explicitly as an additional correction
on equal footing as the nonlocal one.

In practice, 
we first reduce the problem to a low-energy
many-body problem where only the local interactions are
dynamical, but non-local ones are static. Following the
strategy of Hirayama {\it et al.}~\cite{Hirayama}, we treat the non-local dynamical
part of the interactions in a perturbative fashion.
This amounts to 
(a) replacing the non-local dynamical interactions
$[W_{\rm R}^{\rm dyn}(q,\omega)]_{\rm nonlocal}\equiv W_{\rm R}(q,\omega) - \sum_q W_{\rm R}(q,\omega)$ 
by static non-local interactions
$[W_{\rm R}(q,\omega=0)]_{\rm nonlocal}$
and 
(b) treating the frequency-dependent correction 
$W_{\rm R}^{\rm dyn}(q,\omega)$
perturbatively as an additional self-energy
correction. $W_{\rm R}^{\rm dyn}(q,\omega)$ takes on
the form 
defined in Eq.~(\ref{Wrdyn}) or Eq.~(\ref{WdynGW}) 
depending on whether the Hartree-like treatment (denoted
as R=H) or the GW-like treatment (denoted as R=GW) is chosen,
and -- depending on this choice -- leads the to the correction
Eq.~(\ref{DeltaSigmaH}) or Eq.~(\ref{DeltaSigmaGW})
respectively.

Such a perturbative correction can be done in two different
ways: The straightforward option is simple first order
perturbation theory in the difference
$[W_{\rm R}^{\rm dyn}(q,\omega)]_{\rm nonlocal}$.
This part contains the effects of the frequency-dependence of the 
interaction neglected in the effective Hamiltonian formalism 
with $W_{\rm R}(q,\omega=0)$. Again R denotes either the
Hartree-like or GW-like treatment (``H'' or ``GW''). 
We will discuss both
options in the following paragraphs.

\subsubsection{Direct perturbation theory}
\label{directfock}
To first order, the simple perturbative option results in a
correction term
(for simplicity, we drop the frequency summation here)
\begin{eqnarray}
\Delta \Sigma_{\rm L}^{\rm nonlocal}(q)
&=& \sum_{q'}  G^{(1)}(q') [W_{\rm H}^{\rm dyn}(q+q') 
]_{\rm nonlocal} \nonumber \\
&\equiv& \sum_{q'}  G^{(1)}(q') [W_{\rm H}^{\rm dyn}(q+q') 
\nonumber \\
 &-& \sum_q W_{\rm H}^{\rm dyn}(q+q') 
] \nonumber \\
&\simeq& [\sum_{q'}G^{(1)}(q') W_{\rm H}^{\rm dyn}(q+q')
]_{\rm nonlocal}. \nonumber\\
\label{DeltaSigmaLnonlocalPert}
\end{eqnarray}
Here,
\begin{eqnarray}
G^{(1) -1}\equiv G^{(0) -1}+V_{\rm xc}-\Delta \Sigma_{\rm H},
\label{G(1)}
\end{eqnarray}
and $W_{\rm H}^{\rm dyn}(q,\omega)$ is defined in Eq.~(\ref{Wrdyn}).
We stress that the last line of Eq.~(\ref{DeltaSigmaLnonlocalPert})
is not strictly the same as the first line because of the 
nonzero overlap of the single- and two-particle Wannier bases,
as discussed in the Appendix and in Ref.~\onlinecite{sakuma14}.
Nevertheless, as discussed in the Appendix, for 
sufficiently localized basis sets, the difference between the
two previous lines of Eq.~(\ref{DeltaSigmaLnonlocalPert}) 
is tiny and will be neglected hereafter.
In later discussions, we describe this nonlocal part of the 
self-energy in a simplified notation as 
\begin{eqnarray}
\Delta \Sigma_{\rm L}^{\rm nonlocal}(q)
&=&  G^{(1)} [W_{\rm H}^{\rm dyn}(q) 
) ]_{\rm nonlocal} 
\ {\rm or \ equivalently} \nonumber \\
&=&  [G^{(1)} W_{\rm H}^{\rm dyn}(q) 
]_{\rm nonlocal}
\label{DeltaSigmaLnonlocalPert_simple}
\end{eqnarray}

\subsubsection{GW-type perturbation theory}
\label{gwlike}

Alternatively, a more refined perturbation theory
inspired by the GW approximation can be constructed 
for the nonlocal part of Eq.~(\ref{DeltaSigmaGW}) as

\begin{eqnarray}
\Delta \Sigma_{\rm L}^{\rm nonlocal}
&=& 
G_{ll}^{(0)} W^{\rm dyn}_{\rm GW} 
-
G_{ll}^{(0)} 
\left[
W^{\rm dyn}_{\rm GW} 
\right]_{\rm local}.
\label{DeltaSigLnonlocalGW}
\end{eqnarray}
Here, 
$W^{\rm dyn}_{\rm GW}$ as defined in Eq.~(\ref{WdynGW})
corresponds to
the frequency-dependent part of the interaction 
that would be missing if the low-energy part were solved
within the GW approximation.
This justifies to employ the static effective interaction 
$W_{\rm H}(q,\omega=0)=W_{\rm L}(q,\omega\rightarrow \infty)$, 
because $W_{\rm L}(q,\omega)$ is the GW counterpart 
of what will be treated within
the low-energy solver afterwards.

We note that without the subtraction of the local part, this 
correction would correspond to what has been constructed as
$\Delta \Sigma_{\rm L}$ by Hirayama {\it et al.} in Ref.~\onlinecite{Hirayama}.
The local part is not touched here 
since it will be treated nonperturbatively below, following the work
by Casula {\it et al.}~\cite{casula-ZB}. 

Here, $G^{(0)}$ is used in the spirit of a (non-self-consistent)
``one-shot GW" scheme.
If one employs a (partially) self-consistent version of the GW 
scheme, $G^{(0)}$ may be replaced by $G^{(2)}$ defined by 
\begin{eqnarray}
G^{(2) -1}_{ll}\equiv G^{(0) -1}_{ll}+V_{\rm xc}-\Delta \Sigma_{\rm H}-G^{(0)}_{ll}W.
\label{fullGW_G}
\end{eqnarray}
At this level the effective dispersion is renormalized to
\begin{eqnarray} 
T_{\rm eff}^{(1)}(q)&=&[\epsilon_{\rm DFT}-V_{\rm xc} +\Delta \Sigma_{\rm H} (q,\omega =0) \nonumber \\
&+&\Delta \Sigma_{\rm L}(q,\omega =0)]Z_{\rm HL},
\label{T:Sigma_H+DeltaSigma_L}
\end{eqnarray}
where
$Z_{\rm HL}=
(1-
d[\Delta \Sigma_{\rm H} (q,\omega)+\Delta \Sigma_{\rm L}^{\rm nonlocal}(q,\omega)]/d\omega|_{\omega=0}
)^{-1}$
Note that here we have included 
$\Delta \Sigma_{\rm L}^{\rm local}(q,\omega =0)$
as a direct correction, as discussed above.
Together with the nonlocal part
$\Delta \Sigma_{\rm L}^{\rm nonlocal}(q,\omega =0)$
it is thus the full 
$\Delta \Sigma_{\rm L}(q,\omega =0)$
that enters.

\subsection{Intra-d exchange}

The resulting many-body problem with long-range interactions
will have an intra-L-space exchange self-energy contribution of 
Fock form. Also this term takes different forms
depending on whether one places oneself in the perspective
of option (\ref{directfock}) or (\ref{gwlike}) above.

\subsubsection{Direct perturbation theory}
\label{directfock2}

In the first case, the exchange term is the simple Fock
exchange calculated with the static interaction $W_{\rm H}(q,\omega=0)$:
\begin{eqnarray}
\Delta \Sigma^x_{\rm L} = G^{(1)}_{ll} W_{\rm H}(q,\omega=0).
\end{eqnarray}
Once this term has been taken into account, only the correlation
part of the self-energy will have to be calculated within the
low-energy effective model.

However, in most practical many-body calculations, local exchange
contributions will be kept within the low-energy description in the form
of Hund's coupling terms. We therefore prefer to incorporate only
the nonlocal contribution in the one-shot GW as 
\begin{eqnarray}
\Delta \Sigma^{x \ {\rm nonlocal}}_{\rm L} &=& G^{(1)}_{ll} W_{\rm H}(q,\omega=0) \nonumber \\ 
&-& \sum_q [G^{(1)}_{ll} W_{\rm H}(q,\omega=0)]
\end{eqnarray}
into the effective one-body Hamiltonian
while keeping the local one as a many-body term.

An interesting cancellation is observed when the intra-L-space
exchange is combined with the above correction 
$\Delta \Sigma_{\rm L}^{\rm nonlocal}$;
the remaining correction 
\begin{eqnarray}
&& \Delta \Sigma_{\rm L}^{\rm nonlocal} + \Delta \Sigma^{x \ {\rm nonlocal}}_{\rm L} 
\nonumber
\\
&=& [G^{(1)}_{ll} (W_{\rm H}(q,\omega) - W_{\rm H}(q,\omega=0))]_{\rm nonlocal} 
\nonumber
\\
&+& [G^{(1)}_{ll} W_{\rm H}(q,\omega=0)]_{\rm nonlocal}
\nonumber
\\
&=& [G^{(1)}_{ll} W_{\rm H}(q,\omega)]_{\rm nonlocal} 
\label{nlocal_pert}
\end{eqnarray}
reduces to the nonlocal part of a dynamical Fock term, calculated
with the interaction $W_{\rm H}(q,\omega)$, that is, 
the bare interaction within the low-energy space.

\subsubsection{GW-type perturbation theory}
\label{gwlike2}

If, however, the GW-like option is chosen for eliminating the 
frequency-dependence of the nonlocal interactions in the
low-energy subspace (paragraph (\ref{gwlike}) above), the
intra-L-space exchange should accordingly be interpreted as
a screened exchange term.
In practice, this means that again a GW-type expression has
to be adopted:
\begin{eqnarray}
\Delta \Sigma^{x \ {\rm nonlocal}}_{\rm L} = [G^{(0)}_{ll} W_{\rm L}(q,\omega)]_{\rm nonlocal} 
\end{eqnarray}
Combining this term with the above
$\Delta \Sigma_{\rm L}^{\rm nonlocal}$, a similar cancellation as above
is observed:
\begin{eqnarray}
& &\Delta \Sigma_{\rm L}^{\rm nonlocal} + \Delta \Sigma^{x \ {\rm nonlocal}}_{\rm L}
\nonumber
\\ 
&=& [G^{(0)}_{ll} (W(q,\omega) - W_{\rm L}(q,\omega ) )]_{\rm nonlocal}
\nonumber
\\ 
&+& [G^{(0)}_{ll} W_{\rm L}(q,\omega)]_{\rm nonlocal}
\nonumber
\\
&=& [G^{(0)}_{ll} W(q,\omega)]_{\rm nonlocal} 
\label{nlocal_gw}
\end{eqnarray}
The final correction is thus simply the nonlocal part of the
usual GW self-energy
\cite{footnote}.

At this stage, the effective dispersion is renormalized to
\begin{eqnarray}
T_{\rm eff}^{(1)}(q)&=&[\epsilon_{\rm DFT}-V_{\rm xc} +\Delta \Sigma_{\rm H} (q,\omega =0)
\nonumber \\
&+&\Delta \Sigma_{\rm L}^{\rm nonlocal}(q,\omega =0)
\nonumber \\
&+&\Delta \Sigma^{x \ {\rm nonlocal}}_{\rm L}(q,\omega =0)]Z_{\rm HW}.
\label{T:Sigma_H+Sigma_W}
\end{eqnarray}
where 
\begin{eqnarray}
Z_{\rm HW} &=& 
(1-
d[\Delta \Sigma_{\rm H} (q,\omega)+\Delta \Sigma_{\rm L}^{\rm nonlocal}(q,\omega) \nonumber \\
&+& \Delta \Sigma^{x \ {\rm nonlocal}}_{\rm L}(q,\omega)]/d\omega|_{\omega=0}
)^{-1}. 
\end{eqnarray}

\subsection{Frequency-dependence of interactions within low-energy
space: local part}

The remaining problem is one with 
dynamical
local interactions, for which the {\it correlation part} of the
self-energy should be calculated. It can be reduced to a problem with purely
static interactions following Casula {\it et al.}\cite{casula-ZB}: The recipe is to replace
the local dynamical interactions
by static local interactions
while at the same time renormalizing the one-body part of the
problem. 
A subtlety consists however in defining which
dynamical interactions to take.
We again differentiate the two options above.
We also note that the self-energy from local dynamical interaction at zero frequency 
$\Delta\Sigma_{\rm L}^{\rm loc}(\omega=0)$ is already taken into account in Eq.~(\ref{T:Sigma_H+DeltaSigma_L}).

\subsubsection{Direct perturbation theory}
\label{directfock3}

In this case, the additional renormalization factor resulting
from the frequency-dependence of the local interaction
is the one derived in the original work by Casula {\it et al.}:
Indeed, $W^{\rm loc}_{\rm H}(\nu)\equiv \sum_q W_{\rm H}(q,\nu) $,
the effective dynamical interaction in the low-energy subspace
corresponds to what is usually considered as local ``Hubbard $U$'', 
and its frequency-dependence determines the renormalization factor
according to
\begin{eqnarray}
Z_B & = & \exp \left( 1/\pi  \int_0^\infty \!\!\! d\nu
  ~  \textrm{Im} W^{\rm loc}_{\rm H}(\nu) /\nu^2 \right).
\label{Z_B}
\end{eqnarray}

\subsubsection{GW-type perturbation theory}
\label{gwlike3}

The GW-type strategy yields a more involved recipe:
Defining the local part of Eq.~(\ref{WdynGW}), namely
\begin{eqnarray}
[W^{\rm dyn}_{\rm GW} (\omega)]_{\rm local} = 
\sum_q (W - W_{\rm L}) ,
\end{eqnarray}
one can consider that the model to be treated as this stage is
one with an interaction the static part of which is given by
$W_{\rm H}(q,\omega=0)$ 
while its dynamical
part reads $[W_{\rm GW}^{\rm dyn}]_{\rm loc}$. 
(By construction $W^{\rm dyn}_{\rm GW}$ vanishes at zero frequency.)
The corresponding renormalization is given by:
\begin{eqnarray}
Z_B & = & \exp \left( 1/\pi  \int_0^\infty \!\!\! d\nu
  ~  \textrm{Im}  [W^{\rm dyn}_{\rm GW}(\nu)]_{\rm loc} /\nu^2 \right).
\label{Z_B2}
\end{eqnarray}
\section{Summary of the scheme:}
\label{summary}

\noindent
[1] Putting the above steps together, one obtains
the total constrained self-energy 
\begin{eqnarray}
\Delta\Sigma=\Delta\Sigma_{\rm H}+\Delta\Sigma_{\rm L}^{\rm nonlocal}+\Delta\Sigma_{\rm L}^{x \rm{nonlocal}}
\label{DeltaSigma}
\end{eqnarray}
resulting in the following
scheme 
:
\begin{itemize}
\item Calculate the LDA Hamiltonian in the localized basis.
Block-diagonalize it.
\item Calculate the sum of the correction self-energies
above:
\begin{eqnarray}
\Sigma_{\rm corr} (q,\omega)
&=& 
-V_{\rm xc}+\Delta \Sigma.
\end{eqnarray} 
If the simple perturbative strategy is adopted, we employ
$\Sigma_{\rm corr} =\Sigma_{\rm corr}^{\rm Pert}$ (see \ref{directfock}, and \ref{directfock2}):
\begin{eqnarray}
\Sigma_{\rm corr}^{\rm Pert} (q,\omega)
&\equiv &
\nonumber \\ 
-V_{\rm xc}+G^{(0)}_{hh} W_{lhhl}
&+&[G_{ll}^{(1)} W_{\rm H}]_{\rm nonlocal}
\end{eqnarray}
Following the GW-type perturbation theory, it
becomes $\Sigma_{\rm corr}=\Sigma_{\rm corr}^{\rm GW}$(see \ref{gwlike}, and \ref{gwlike2}):
\begin{eqnarray}
\Sigma_{\rm corr}^{\rm GW} (q,\omega)
&\equiv & 
\nonumber \\
-V_{\rm xc}+G^{(0)}_{hh} W_{lhhl}
&+&[G_{ll}^{(0)} W]_{\rm nonlocal}.
\end{eqnarray}
This self-energy can be linearized, e.g. around the Fermi
level, giving rise to a static correction
$\Sigma_{\rm corr} (q,\omega=0)$
and a $Z$-factor corresponding to its (linearized)
frequency-dependence:
\begin{eqnarray}
Z_{\rm corr} = \frac{1}{1- \frac{\partial \Sigma_{\rm corr} (q,\omega)}
{\partial \omega}|_{\omega=0}} .
\end{eqnarray}
\end{itemize}
[2] The effect of the local dynamical interaction is taken into account as follows:
\begin{itemize}
\item Calculate the renormalization factor $Z_B$ arising from the local part of the 
frequency dependence in the screened interaction (see \ref{directfock3}, and \ref{gwlike3}).
\end{itemize}
[3]  The effective low-energy Hamiltonian is eventually given by
\begin{eqnarray}
H_{\rm eff}^{(1)}&=&\sum_q T_{\rm eff}^{(1)}(q)c^{\dagger}_qc_q \nonumber \\
&+&\sum_{q,k,p} W_{r}(q,0)c^{\dagger}_k c_{k+q}c^{\dagger}_p c_{p-q}.
\label{first-Heff}
\end{eqnarray}

with 
\begin{eqnarray}
T_{\rm eff}^{(1)}(q)=[\epsilon_{\rm DFT}+\Sigma_{\rm corr} (q,\omega=0)]Z_{\rm corr}Z_B.
\label{Teff_final}
\end{eqnarray}

\noindent
[4] Solve the many-body problem with the one-body Hamiltonian
from the preceding step and the static non-local interactions
$W_{\rm H}(q,\omega=0)$.
Take care of removing the Hartree contribution from the solution 
of the many-body problem, in order to avoid double counting
with the initial single-particle Hamiltonian where the Hartree
potential was included in LDA.
This can be done by following the procedure in Ref.~\onlinecite{Hirayama}, where the Hartree solution of the low-energy effective
model is subtracted before solving by the low-energy solver.
This can also be done as in LDA+U or LDA+DMFT techniques, by
calculating the Hartree solution of the effective low-energy
model.

\section{Variants of the low-energy model}
\label{variants}

The above discussion has focused on the construction of an
effective low-energy many-body problem with static and local 
interactions, for which only the correlation part of the 
self-energy needs to be calculated from the many-body solver.
Alternatively, if one uses a many-body solver that can fully
handle long-range interactions,
a variant of the above scheme can be 
envisioned. 
Another variant can be used if one wishes to construct a low-energy model with purely.
local interactions only.
A different variant is useful if one wishes to update the single-particle
part of the Hamiltonian, after an improved estimate for the 
electronic density is available after the many-body calculation.
Those are the subject of the following paragraphs.

\subsection{Low-energy model with long-range interactions}

If the aim is the construction of a low-energy effective model
with fully long-range Coulomb interactions, the treatment of
the intra-L-space exchange term can be omitted at the level
of the construction of the model.
The remaining corrective self-energy reads:
\begin{eqnarray}
\Sigma_{\rm corr} (q,\omega )
&=& 
\Delta \Sigma_{\rm H} + \Delta \Sigma_{\rm L}^{\rm nonlocal}
\nonumber
\\
&=& 
G^{(0)}_{hh} W_{lhhl}
+
G^{(0)}_{ll} 
W^{\rm dyn}_{\rm H}
\nonumber
\\
&-&
\left[
G^{(0)}_{ll} 
W_{\rm H}^{\rm dyn}
\right]_{\rm local}
\end{eqnarray}
where the option of
direct perturbation theory
$W^{\rm dyn}_{\rm H}$  defined in Eq.~(\ref{Wrdyn}) should be
replaced by 
$W^{\rm dyn}_{\rm GW}$ defined in Eq.~(\ref{WdynGW}) for the option of
the GW-type perturbation theory.

We note however, that the reduction to a static model using
the Casula procedure involves in this case an additional approximation:
indeed, strictly speaking, the Casula procedure modifies the
non-density-density terms of the interactions, by dressing the
creation and annihilation operators with exponential weight
factors. This can be seen as follows: 
the Casula procedure is based on a Lang-Firsov transformation
\cite{lang_firsov}, replacing
the original fermionic operators  $c_{i\sigma}$, at $i$th site with spin $\sigma (= \uparrow$ or $\downarrow )$, by polaronic
operators $d_{i \sigma}$, thus eliminating the explicit dependence
on the bosonic operators $b_i$ that describe the screening
degrees of freedom:
$d_{i \sigma} =  \exp(\lambda (b_i - b_i^{\dagger})) c_{i \sigma}$.
While the exponentials drop out for density-density terms,
since $d_{i \sigma}^{\dagger}  d_{i \sigma} =c_{i \sigma}^{\dagger}  c_{i \sigma}$,
this is not true for more general interactions.
In principle, the corresponding correction factors can be
worked out by straightforward operator algebra. For
simplicity, we will however disregard here this complication,
assuming e.g. that any long-range interactions are of pure
density-density type.

\subsection{Effective model with local Hubbard interactions only}

The solution of the final many-body problem with
nonlocal interactions may in principle be done within
many-body solvers suitable for non-local
interactions such as various Monte Carlo methods including the variational Monte Carlo~\cite{Tahara,Gros}.
Alternatively, extended DMFT (EDMFT)~\cite{sun,ayral} can be viewed as a means to determine
an effective local interaction that best represents the effects
of the initial long-range interactions, and can be
considered as a technique to ``backfold'' long-range interactions
into effective local ones.

Very generally, from the above discussion it becomes obvious
that the construction of the one-particle part of the Hamiltonian
will depend on which interaction terms will be included in the
many-body calculation, while physical properties obtained after 
solving the low-energy problem are expected to be insensitive to the choice. 
In the next section, we will see how the derived effective models 
behave in the case of the simple oxide SrVO$_3$.

\subsection{Update of single-particle Hamiltonian}
In some cases, many-body effects substantially change the charge
distribution in the low-energy subspace as compared to the LDA
one. Such redistributions of charge can for example happen between
different orbitals in multi-orbital systems, and have actually
been observed e.g. in titanium oxides~\cite{pavarini-d1},
BaVS$_3$~\cite{bavs3} or
iron-based superconductors \cite{aichhorn10, Hirayama2015}.
If this happens, one might want to update the
starting Hamiltonian and GW self-energies, based on the
new charge density rather than the converged LDA one,
analogously to what is done in the DFT+DMFT calculations
\cite{Pu-Savrasov, ce2o3}.
This effect can induce non-negligible
corrections to the relative orbital levels.

Technically, the resulting self-consistency loop is analogous
to what has been discussed in detail in the DMFT literature
\cite{lechermann-d1}, in particular concerning the way the density
is recalculated in the continuum after the solution of the
effective model.

\section{Results}
\label{results}

The ternary 3d$^1$ transition metal oxide SrVO$_3$ has become
one of the ``drosophila compounds'' of correlated systems.
It is a correlated metal
that has been characterized using various experimental 
\cite{Onoda, inoue, maiti, maiti2, sekiyama, takizawa, yoshida2010}.
and theoretical techniques, see e.g.
\cite{nekrasov, anisimov, pavarini-d1, liebsch, lechermann-d1, trimarchi, aichhorn09, karolak, lee, held-optics, tomczak-gwdmft, gatti-guzzo, sakuma14, TomczakPRB2014}.
A review of most of the available experimental and theoretical 
data has been given recently in Ref.~\onlinecite{TomczakPRB2014}.
SrVO$_3$ displays Fermi liquid behavior up to
remarkably high temperatures of the order of $200$ K \cite{Onoda, inoue},
with a moderate mass enhancement of the order 
of $2$ \cite{yoshida2010, vandermarel}.
Detailed spectroscopic investigations have made it an
ideal test compound for modern many-body calculations, and 
more and more refined dynamical mean field-based studies 
are available.
The majority of studies so far have focused on the t$_{2g}$-manifold
which forms the states close to the Fermi level, and those will
also be the focus in the present investigations. 
Note however that this restriction quite severely limits the
range of validity of the low-energy description. Indeed, as
shown recently~\cite{tomczak-gwdmft, TomczakPRB2014} at
energies of about 2 to 3 eV above the Fermi level
the spectral properties are largely determined by the e$_g$
states. This has in particular led to a reinterpretation of
an inverse photoemission feature at about 2.5 eV that was 
frequently interpreted as an upper Hubbard band of t$_{2g}$
character in the earlier literature.
Here, we use the compound only as an illustration of 
the principles of constructing effective low-energy models,
without being concerned with a description of spectra beyond
the pure t$_{2g}$ part.

\begin{figure}[h]
\centering 
\includegraphics[clip,width=0.45\textwidth ]{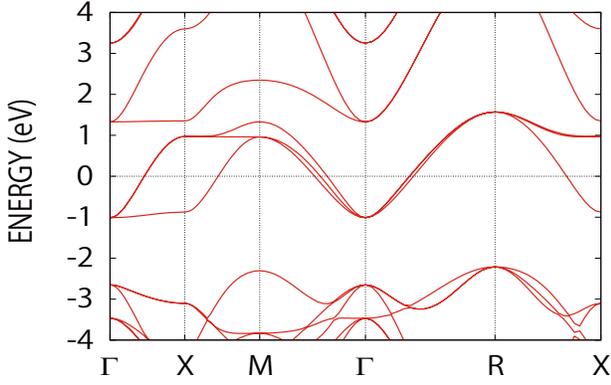} 
\caption{(Color online)
Kohn-Sham band structure of SrVO$_3$ in the LDA.
The energy is measured from the Fermi level.
}
\label{lda}
\end{figure}

The LDA band structure of SrVO$_3$ is shown in Fig.~\ref{lda}.
One clearly distinguishes the three-fold degenerate manifold
of t$_{2g}$ bands close to the Fermi level (chosen as the
zero of energy), followed in the unoccupied part of the spectrum
by the two e$_g$ bands. At about -2.5 eV the filled O-2p derived
bands are visible.

\begin{figure}[h]
\centering 
\includegraphics[clip,width=0.45\textwidth ]{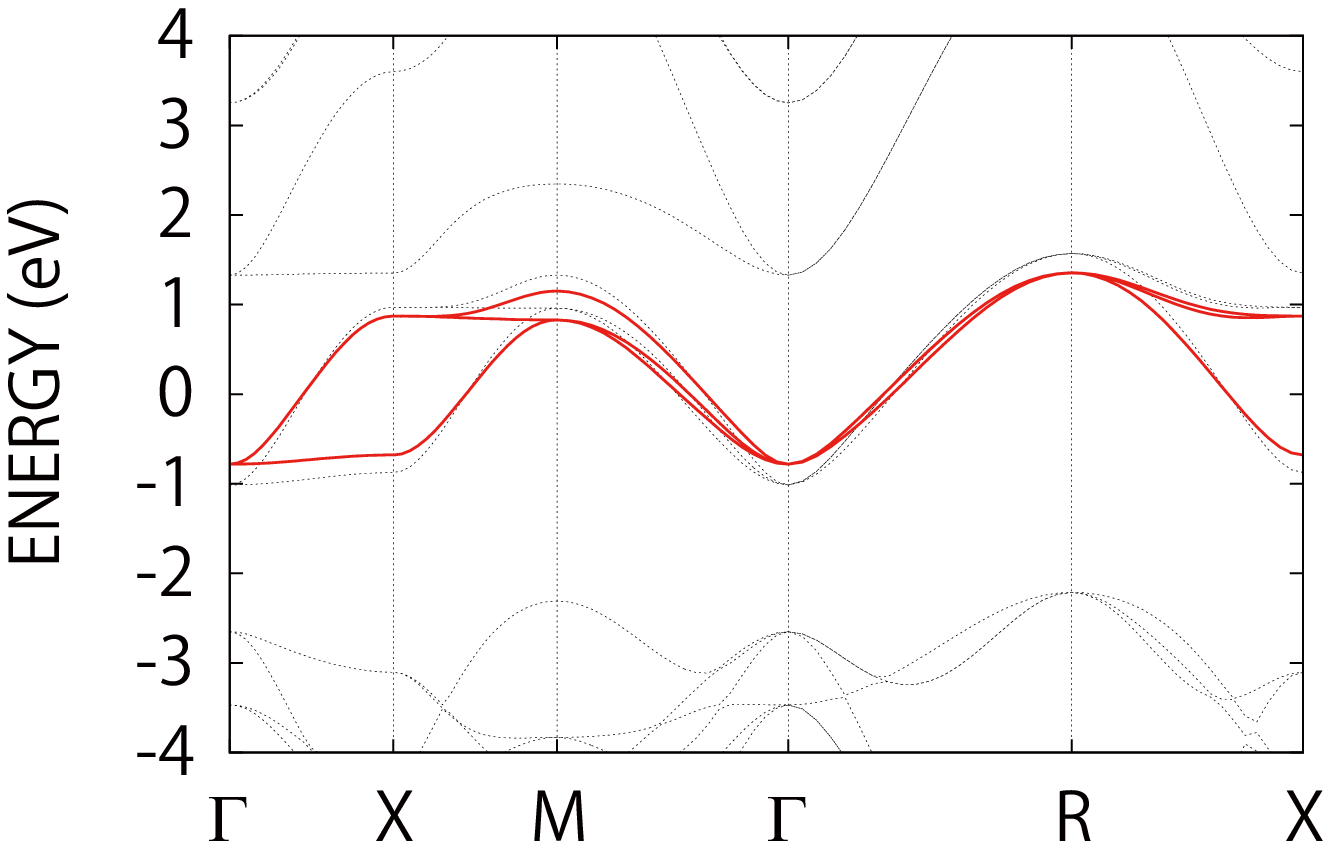} 
\caption{(Color online)
Band structure  of SrVO$_3$ in the one-shot GW approximation.
For comparison, the LDA band structure is also given
(black dotted line).
The energy is measured from the Fermi level.
}
\label{fullGW}
\end{figure}

When a standard GW calculation is performed, see Fig.~\ref{fullGW},
the bandwidth of the t$_{2g}$-states is reduced from 2.5 eV to
2.1 eV, while the overall shape of the dispersion remains similar
to the LDA one.
This is in agreement with previous GW calculations in the literature
\cite{tomczak-gwdmft, TomczakPRB2014, gatti-guzzo, sakuma14}.

The calculation is based on the full-potential linear muffin-tin orbitals (FP-LMTO) implementation.
8$\times$8$\times$8 \bm{k}-mesh is employed in both the DFT/LDA and the GW calculations.
The calculational details are the same as in Ref.~\onlinecite{Hirayama}.

We now turn to a discussion of the band structure corresponding to 
the Hamiltonians to be used as input for subsequent many-body 
calculations for this material.
We will proceed step by step to analyze the influence of the various
correction terms, with respect to the DFT starting point
(which will be overlaid to the respective band structures).

\begin{figure}[h]
\centering 
\includegraphics[clip,width=0.45\textwidth ]{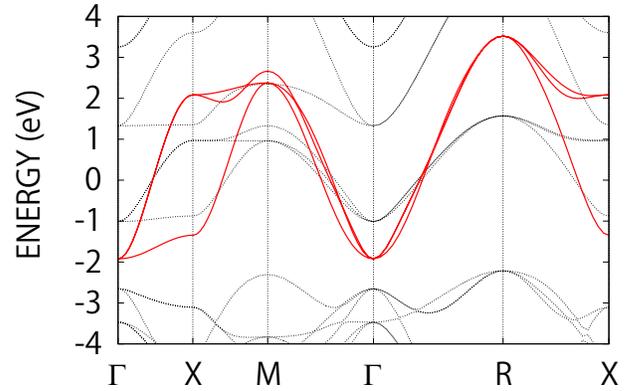} 
\caption{(Color online)
Band structure of SrVO$_3$ corresponding to the LDA Hamiltonian
from which the LDA exchange-correlation potential has been subtracted.
For comparison, the LDA band structure is also given
(black dotted line).
The energy is measured from the Fermi level.
}
\label{lda-vxc}
\end{figure}

\begin{figure}[h]
\centering 
\includegraphics[clip,width=0.45\textwidth ]{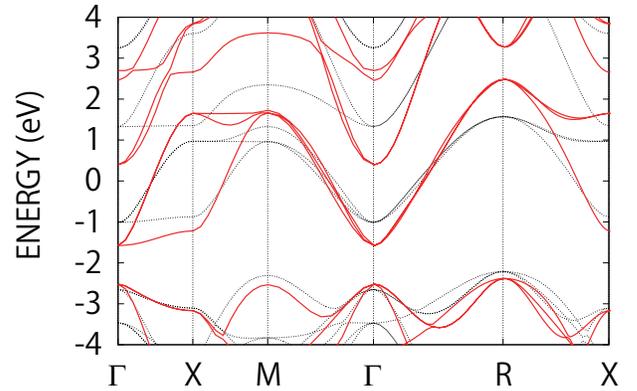} 
\caption{(Color online)
Band structure  of SrVO$_3$ in the Hartree approximation.
For comparison, the LDA band structure is also given
(black dotted line).
The energy is measured from the Fermi level.
}
\label{hartree}
\end{figure} 

Figure~\ref{lda-vxc} 
displays the band structure corresponding to
the LDA Hamiltonian from which the 
LDA exchange-correlation
potential has been subtracted,  
that is, the eigenvalues of $H_{\rm LDA}-V_{\rm xc}$ where
these operators are evaluated for the self-consistent LDA
density.
The subtraction of $V_{\rm xc}$ widens the band structure from 
the LDA bandwidth of 2.5 eV 
to more than twice this value: the new band
width is 5.4 eV.
This indicates that the exchange-correlation potential $V_{\rm xc}$ 
of the LDA is responsible for a substantial amount of 
the band-narrowing effect
arising from electronic correlations, stressing the need 
to subtract the double counting in a consistent manner. 
Figure~\ref{lda-vxc} is regarded as the starting point for the
following considerations.

For simple materials (see e.g. the calculations on Li in Ref.~\onlinecite{Lithium})
it has been noted in the literature that the Hartree band structure
is close to the DFT one. This raises the question of the origin of the
substantial band widening in the present case. Indeed, the present
calculation shown in Fig.~\ref{lda-vxc} differs from a Hartree calculation
only 
by the fact that Fig.~\ref{lda-vxc} is evaluated for the converged LDA density.
We have performed a test calculation where we plot the Hartree band
structure calculated for the converged Hartree density. The result
is shown in Fig.~\ref{hartree}. As seen from this plot, while the
band is not fully as wide as in Fig.~\ref{lda-vxc}, a substantial 
widening is already present at this stage.

\begin{figure}[h]
\centering 
\includegraphics[clip,width=0.45\textwidth ]{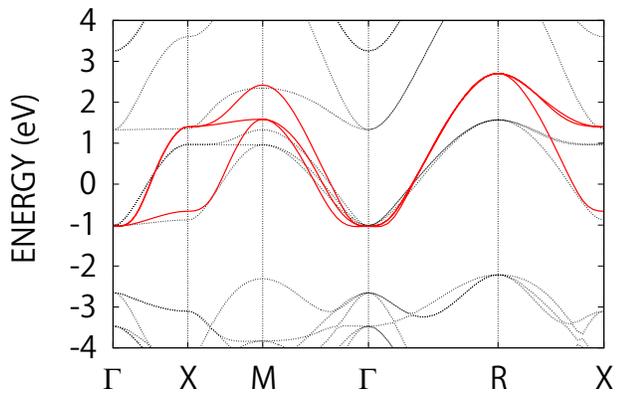} 
\caption{(Color online)
Band structure  of $H_{\rm LDA}-V_{\rm xc}+\Delta \Sigma_{\rm H}(\omega=0)$ of SrVO$_3$.
For comparison, the LDA band structure is also given
(black dotted line).
The energy is measured from the Fermi level.
}
\label{lda-vxc-sigH(0)}
\end{figure} 

Starting from the Hamiltonian without V$_{xc}$ (see the 
dispersion in Fig.~\ref{lda-vxc}), we first take into account 
the static part of the corrective self-energy $\Delta \Sigma_{\rm H}$:
In Fig.~\ref{lda-vxc-sigH(0)}, we plot the dispersion corresponding to
$H_{\rm LDA}-V_{\rm xc}+\Delta \Sigma_{\rm H}(\omega=0)$. 
Compared to the dispersion of Fig.~\ref{lda-vxc}, the overall band structure 
is narrowed to 3.7 eV. This value is, however, still considerably larger
than the LDA bandwidth (Fig.~\ref{lda}). Although
 the electronic correlations coupling the H- and 
the L space are included in Fig.~\ref{lda-vxc-sigH(0)}, and narrow the band with respect to the case
where V$_{xc}$ is taken out, correlations within
the L space are not included. 
The LDA, on the other hand, at least partially
includes correlations within the L space, 
and these are very effective in narrowing the band.
Interestingly, the bottom of the occupied band is quite
exactly at the LDA value, and the remaining widening is
purely in the unoccupied part.

\begin{figure*}[htb]
\begin{center} 
\includegraphics[width=0.9\textwidth ]{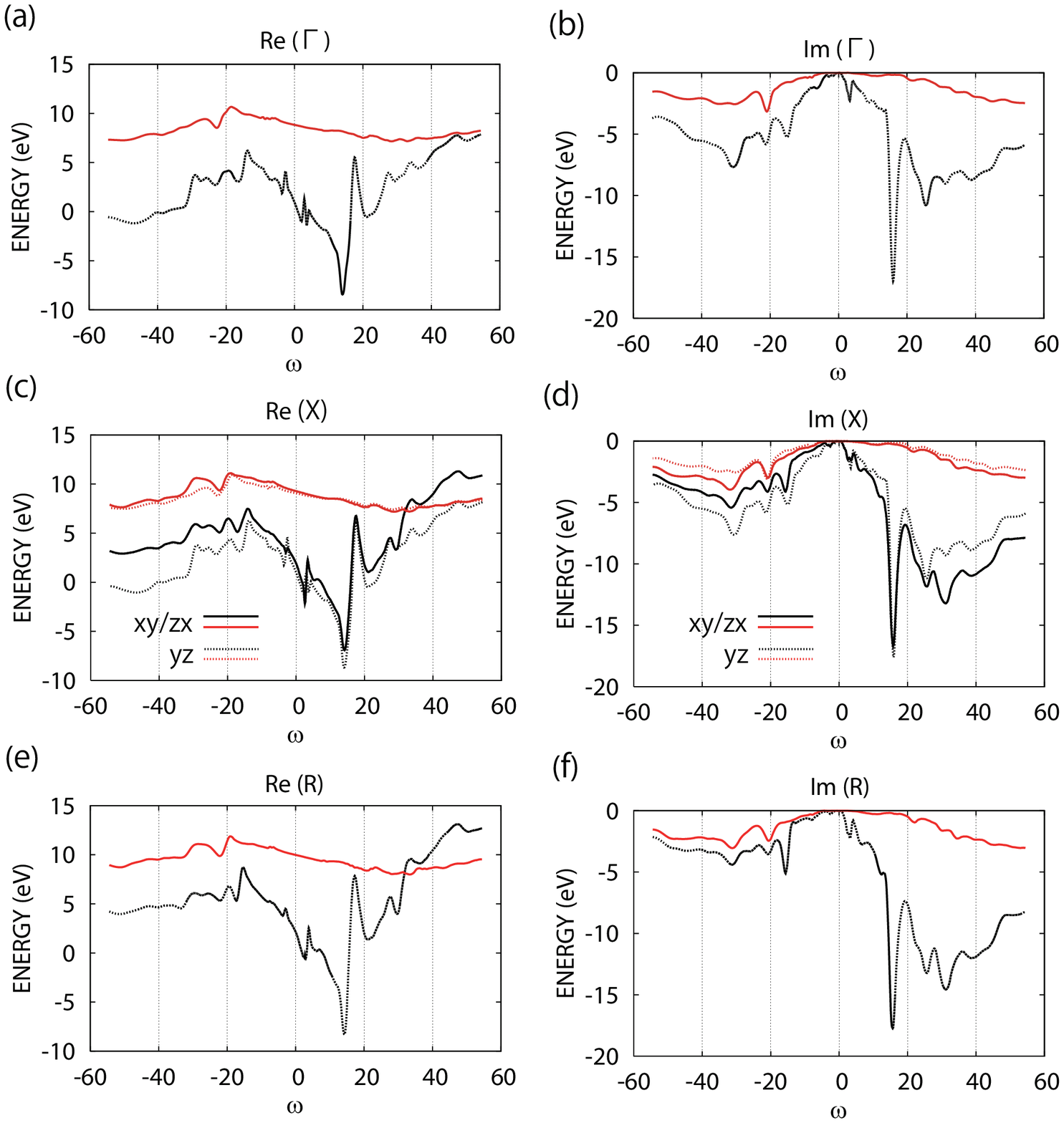} 
\end{center} 
\caption{(Color online)
Frequency dependence of $-V_{xc}+\Delta \Sigma_{\rm H}$ for SrVO$_3$.
(a) and (b) are the real and imaginary parts at the $\Gamma$ point, 
respectively,
and (c) and (d) ((e) and (f)) are those at $X$ ($R$).
For comparison, the full GW self-energies are also given
(black dotted line).
}
\label{sigH}
\end{figure*} 
\begin{figure}[h]
\centering 
\includegraphics[clip,width=0.45\textwidth ]{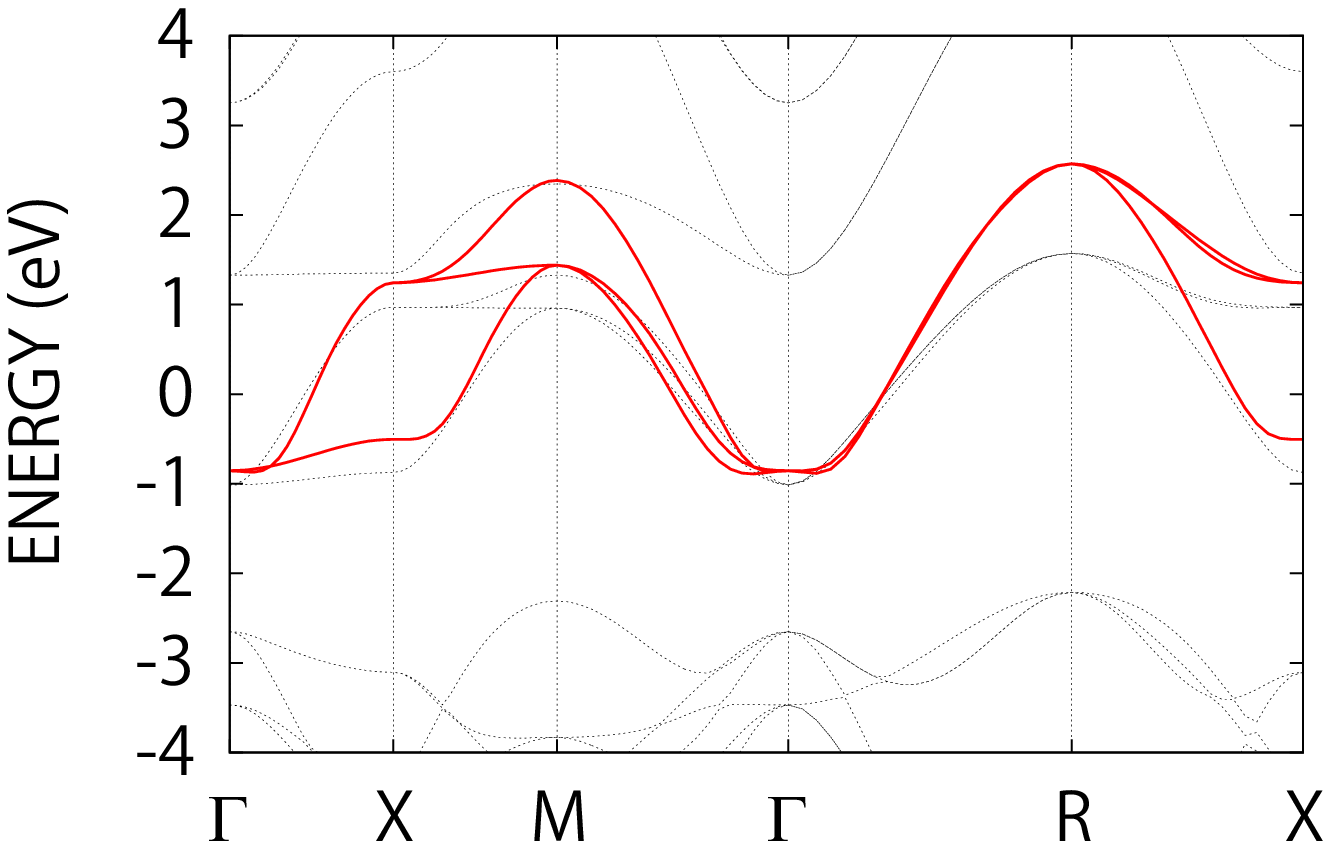} 
\caption{(Color online)
Band structure  of $Z_{\rm H}[H_{\rm LDA}-V_{\rm xc}+\Delta\Sigma_{\rm H}(\omega=0)]$ of SrVO$_3$.
For comparison, the LDA band structure is also given
(black dotted line).
The energy is measured from the Fermi level.
}
\label{ZH_SigmaH}
\end{figure} 

To go further and in particular to analyze the dynamic
behavior of $\Delta \Sigma_{\rm H}$ 
we plot in Fig.~\ref{sigH} 
the self-energy corrections $\Delta \Sigma_{\rm H}$ (with $\Sigma$ (see Eq.~(\ref{full-self-energy}))
for comparison) for the real and imaginary parts
at several representative choices of momenta.
The frequency dependence is smooth around the Fermi level.
In particular, in contrast to the full self-energy,
there are no poles in $\Delta \Sigma_{\rm H}$ 
in the low-energy region, thanks to the
exclusion of fluctuations within the L-space.
The frequency dependence of the real part indicates that the 
linearization 
$\displaystyle{\Delta \Sigma_{\rm H}(\omega) \sim \Delta \Sigma_{\rm H}(\omega=0)+[d\Delta \Sigma_{\rm H}(\omega)/d\omega]_{\omega=0}\omega }$
offers a reasonably good approximation. 
In fact, the behavior of the constrained self-energy
is much closer to linearity than that of 
the full GW self-energy $\Sigma$.
This is easily understood by the fact that the low-energy excitations are excluded in the present constrained self-energy $\Delta\Sigma_{\rm H}$ analogously to an insulator, thus eliminating the strong frequency dependence of the constrained self-energy near the Fermi level. This is one of the consequences of the 
controllability and an advantage of the present MACE scheme, as 
mentioned in the introduction.
In the following discussion, we employ this linearized approximation.
Then the renormalization factor $Z_{\rm H}$ defined by Eq.~(\ref{ZH})
is interpreted as that from the contribution of the H-space. 
After 
this renormalization factor is taken into account,
the effective Hamiltonian is given by $Z_{\rm H}[H_{\rm LDA}-V_{\rm xc}+\Delta\Sigma_{\rm H}(\omega=0)]$ (see Fig.~\ref{ZH_SigmaH}). 
Interestingly the renormalization factor $Z_{\rm H}$=0.92 stays
close to one, so that
the bandwidth is only slightly reduced (to
$3.4$ eV instead of the $3.7$ eV above). 
We also note that $Z_{\rm H}$ has weak momentum dependence
as we reveal in the following.

\begin{figure*}[htb]
\begin{center} 
\includegraphics[width=0.9\textwidth ]{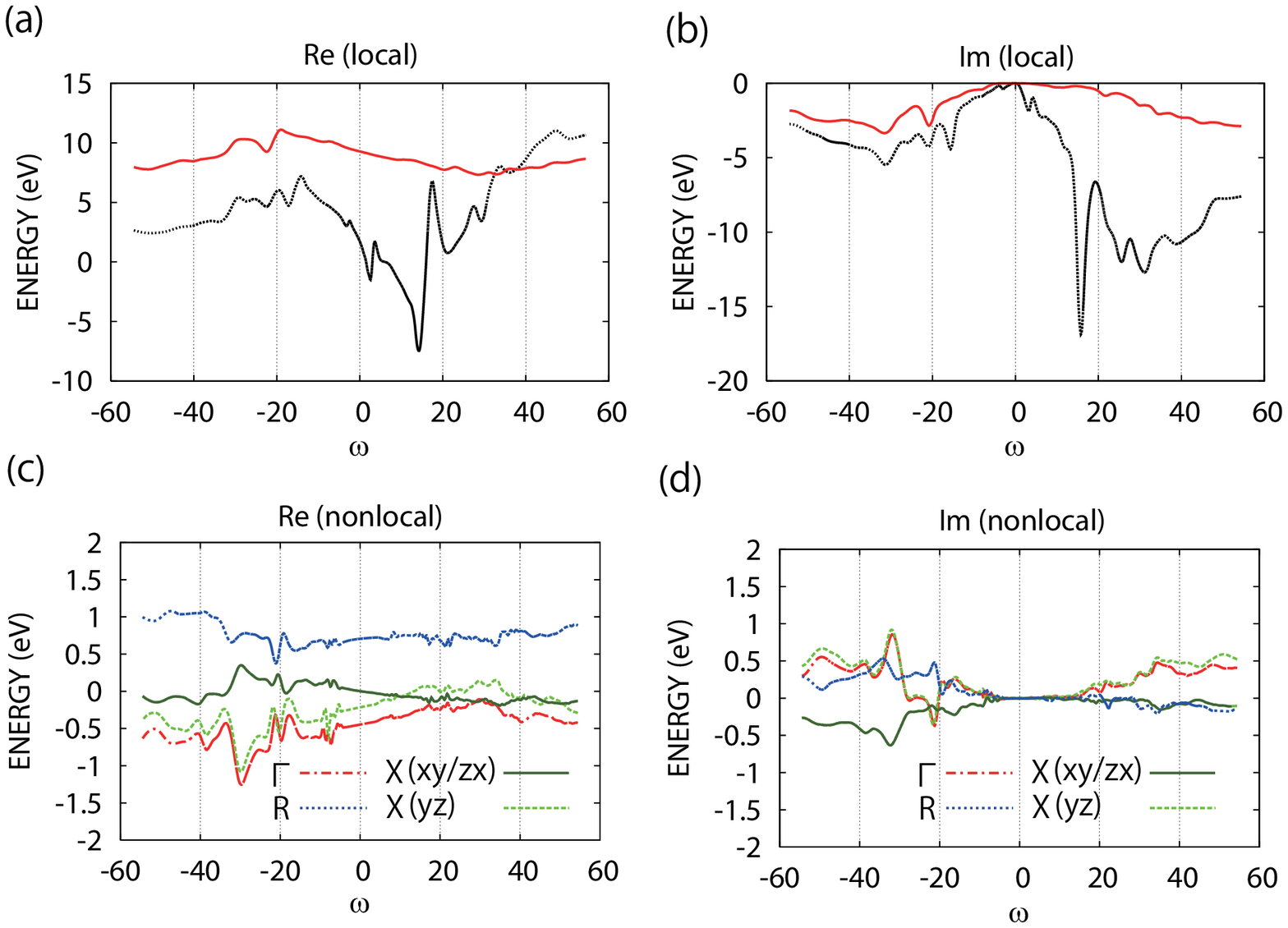} 
\end{center} 
\caption{(Color online)
{\bf minor modification}
Frequency dependence of $-V_{\rm xc}+\Delta \Sigma_{\rm H}$ of SrVO$_3$.
(a) and (b) are the real and imaginary parts of the local
component, respectively.
The black dotted line is $\Sigma$ defined in Eq.~(\ref{full-self-energy}) in the full GW approximation for comparison.
(c) and (d) are the real and imaginary 
parts at several $k$ points.
}
\label{sigH_nonlocal}
\end{figure*} 

Figure~\ref{sigH_nonlocal} 
shows the local and nonlocal parts of $\Delta \Sigma_{\rm H}$.
The frequency 
dependence
of the local part (Figs.~\ref{sigH_nonlocal} (a)(b)) 
is quite similar 
to the  frequency-dependent self-energies at various momenta
in Fig.~\ref{sigH}.
One immediately reads off an interesting property, which is
akin to what has been found for the full GW self-energy in
Ref.~\onlinecite{Tomczak-PRL2012}, namely that the nonlocal part of 
$\Delta \Sigma_{\rm H}$ shows only weak frequency dependence 
as shown in Figs.~\ref{sigH_nonlocal} (c)(d). 
This explains why $Z_{\rm H}$ 
is only weakly momentum-dependent.
On the other hand, $\Delta \Sigma_{\rm H}^{\rm nonlocal} (0)$ at $R$ is $\sim 1$ eV larger than 
at $\Gamma$, which causes the band widening effect.
 $\Delta \Sigma_{\rm H}$
can thus to a good approximation be decomposed into a
frequency-dependent local part and a static nonlocal one:
$\Delta \Sigma_{\rm H}(k,\omega) = \Delta \Sigma^{\rm local}_{\rm H}(\omega) + \Delta \Sigma^{\rm nonlocal}_{\rm H}(k)$,
with $ \Delta \Sigma_{\rm H}^{\rm local}(\omega) = \sum_q  \Delta \Sigma_{\rm H}(q,\omega)$

Since in the simple case of SrVO$_3$ where the t$_{2g}$ states are
degenerate, static local operators are scalar and are compensated 
by a chemical potential shift such that the correct particle
number is obtained. The above band structure corresponding to
$H_{\rm LDA} - V_{\rm xc} + \Delta \Sigma_{\rm H}(k, \omega=0)$ 
is thus identical to that of
$H_{\rm LDA} - V_{\rm xc} + \Delta \Sigma^{\rm nonlocal}_{\rm H}(k)$.
The local dynamical part of $\Delta \Sigma_{\rm H}$ then results in a 
narrowing of this band structure by a factor $Z_{\rm H} = 0.92$.

\begin{figure}[h]
\centering 
\includegraphics[clip,width=0.45\textwidth ]{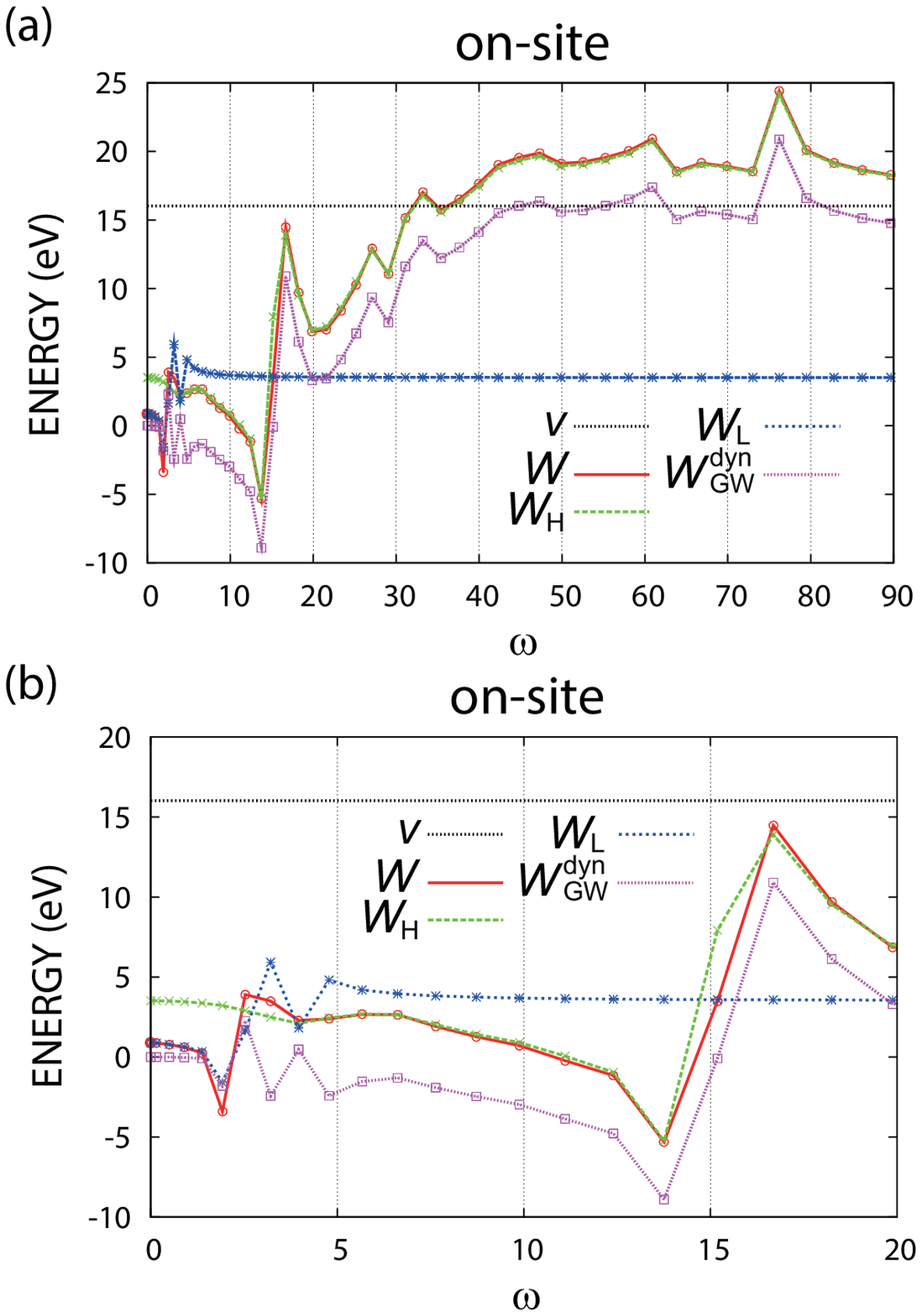} 
\caption{(Color online)
(a) Frequency dependence of the on-site effective interaction.
Panel (b) is a low-energy zoom of panel (a). 
}
\label{W}
\end{figure} 

Before turning to a discussion of the low-energy correction for the GW treatment
$\Delta \Sigma_{\rm L} =G^{(0)}(W-W_{\rm L})$, we analyze the effective
interaction $W_{\rm L}$ shown in Fig.~\ref{W} 
in comparison to $W$ and $W_{\rm H}$.
The fully screened Coulomb interaction $W$ displays the familiar
shape of an interaction that is strongly screened at low energies
(with a value of $0.88$ eV at $\omega=0$), while retrieving the
value of the bare Coulomb interaction $v$ ($\sim 16.0$ eV) 
at high energy.
The crossover from the bare to the  
screened values takes place at the plasma energy of 
about $15$ eV. This behavior has been discussed before
\cite{Review_SB2014}; we note in particular that the plasma 
frequency is known from electron energy loss spectroscopy 
measurements of the related SrTiO$_3$ compound \cite{vandermarel,kohiki}.
The partially screened interaction
$W_{\rm H}$ constructed within cRPA
converges to $W$ at high energies, but displays
weaker screening effects at low energies (with a value of $3.5$ eV at $\omega=0$),
since intra-t$_{2g}$ screening channels are excluded.
As was already anticipated in Sec.~\ref{Outline},  $W_{\rm L}$ can be interpreted as the screened interaction of 
a low-energy model where a static interaction of value $W_{\rm H}(\omega=0)$
has been imposed as the bare interaction. Since now {\it only} intra-t$_{2g}$
screening channels are active, screening takes place only at low
energies, where the scale is given by the t$_{2g}$ bandwidth.
Also plotted is the difference $W_{\rm GW}^{\rm dyn}=W-W_{d}$: except at low energies
where the t$_{2g}$ screening channels come into play, its frequency
dependence is essentially given by that of $W$, while the
high-energy limit is the bare Coulomb interaction $v$ reduced
by $W_{\rm H}(0)$.

\begin{figure*}[htb]
\begin{center} 
\includegraphics[width=0.9\textwidth ]{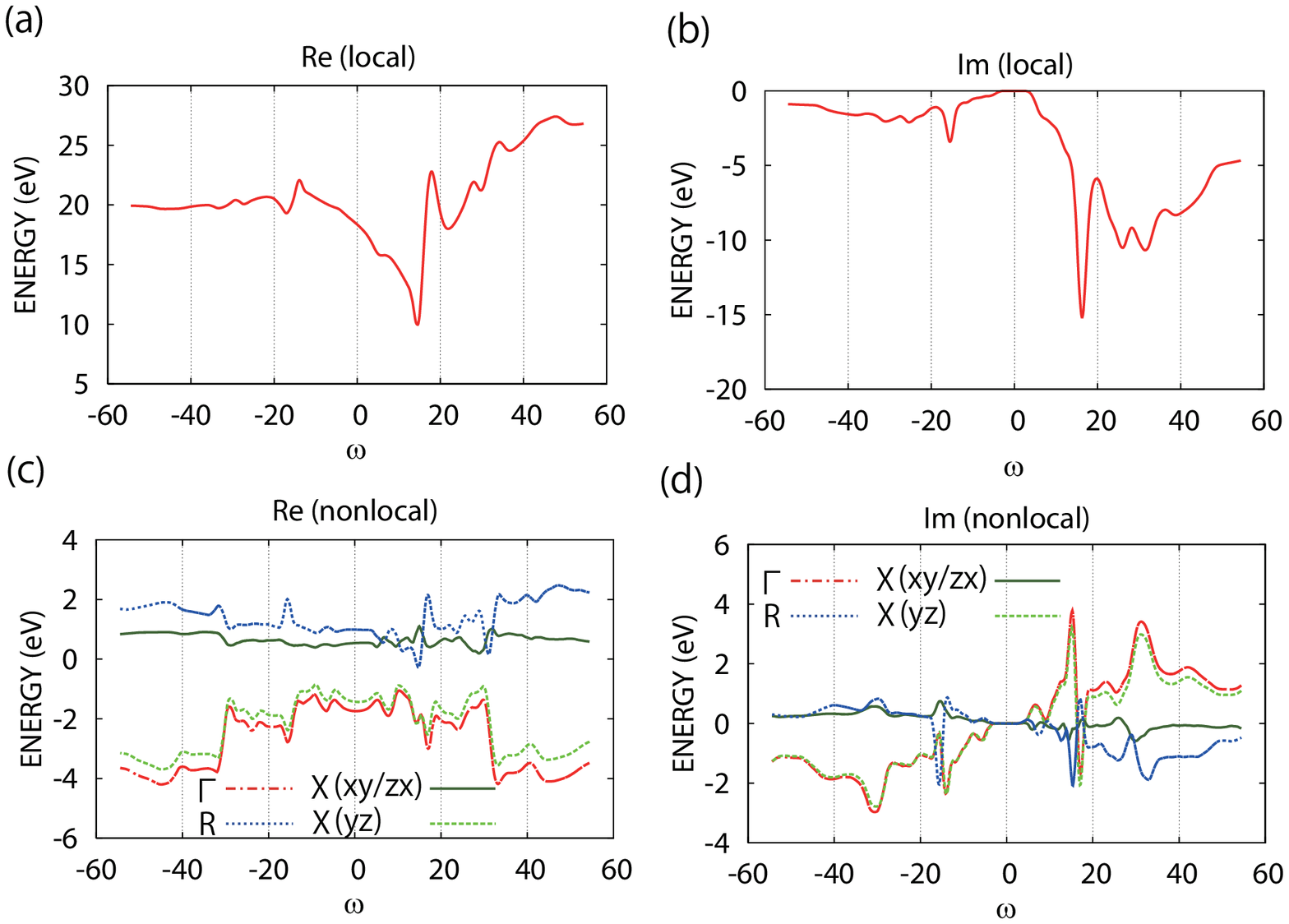} 
\end{center} 
\caption{(Color online)
Frequency dependence of $-V_{\rm xc}+ G^{(1)}W_{\rm H}^{\rm dyn}$ of SrVO$_3$.
(a) and (b) are the real and imaginary local parts, respectively.
(c) and (d) are the real and imaginary 
parts at several $k$ points.
}
\label{DeltaSigma_L_Pert}
\end{figure*} 
\begin{figure*}[htb]
\begin{center} 
\includegraphics[width=0.9\textwidth ]{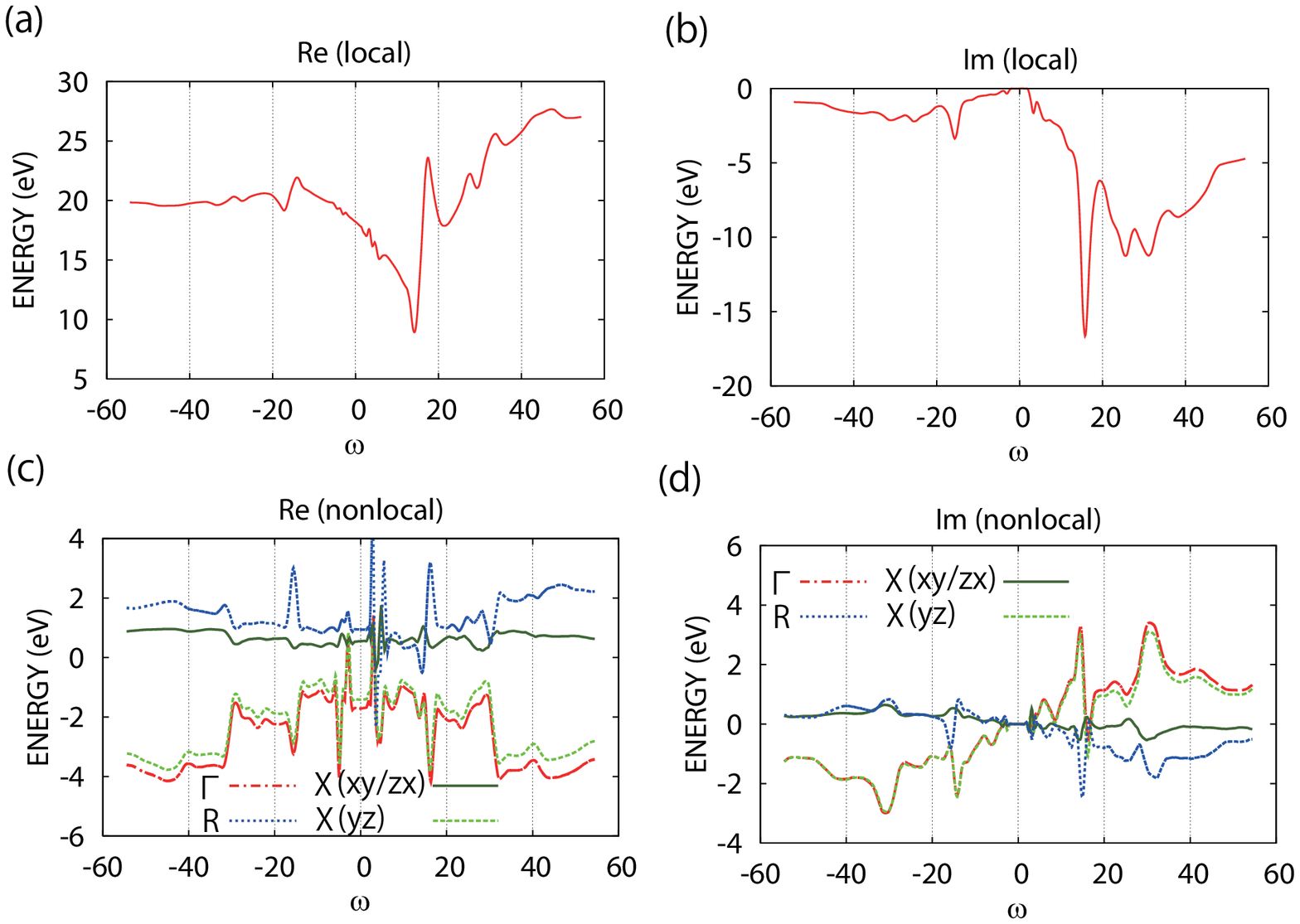} 
\end{center} 
\caption{(Color online)
Frequency dependence of $-V_{\rm xc}+ G^{(0)}W_{\rm GW}^{\rm dyn}$ of SrVO$_3$.
(a) and (b) are the real and imaginary local parts, respectively.
(c) and (d) are the real and imaginary 
parts at several $k$ points.
}
\label{DeltaSigma_L_GW}
\end{figure*} 

The frequency dependence of the real and imaginary parts of the low-energy self-energy $\Delta \Sigma_{\rm L} = G^{(1)}W_{\rm H}^{\rm dyn}$ for the direct perturbative treatment is illustrated in Fig.~\ref{DeltaSigma_L_Pert}. 
$\Delta \Sigma_{\rm L} = G^{(0)}W_{\rm GW}^{\rm dyn}$ used
for the GW treatment is shown in Fig.~\ref{DeltaSigma_L_GW}
for several choices of momenta.
\\
The results for the perturbative and the GW treatment are nearly 
identical.
The renormalization factor of $\Delta \Sigma_{\rm L}^{\rm nonlocal}$ 
(and $\Delta \Sigma_{\rm L}$) is $\sim 0.77$.
The zero-frequency shift
${\rm Re}\Delta \Sigma_{\rm L}^{\rm nonlocal} (0)$ at $R$ is $\sim 3$ eV 
larger than 
at $\Gamma$.
Again, one sees how $\Sigma_{\rm L}$ separates into local dynamical
and nonlocal static parts: 
$\Delta \Sigma_{\rm L} = \Delta \Sigma^{\rm nonlocal}_{\rm L}(k) + 
\Delta \Sigma^{\rm local}_{\rm L}(\omega)$.
This is expected since such a separation has been found within
the full GW calculation \cite{TomczakPRB2014}, and is even
more plausible for a GW treatment within the t$_{2g}$ subspace.

\begin{figure}[h]
\centering 
\includegraphics[clip,width=0.45\textwidth ]{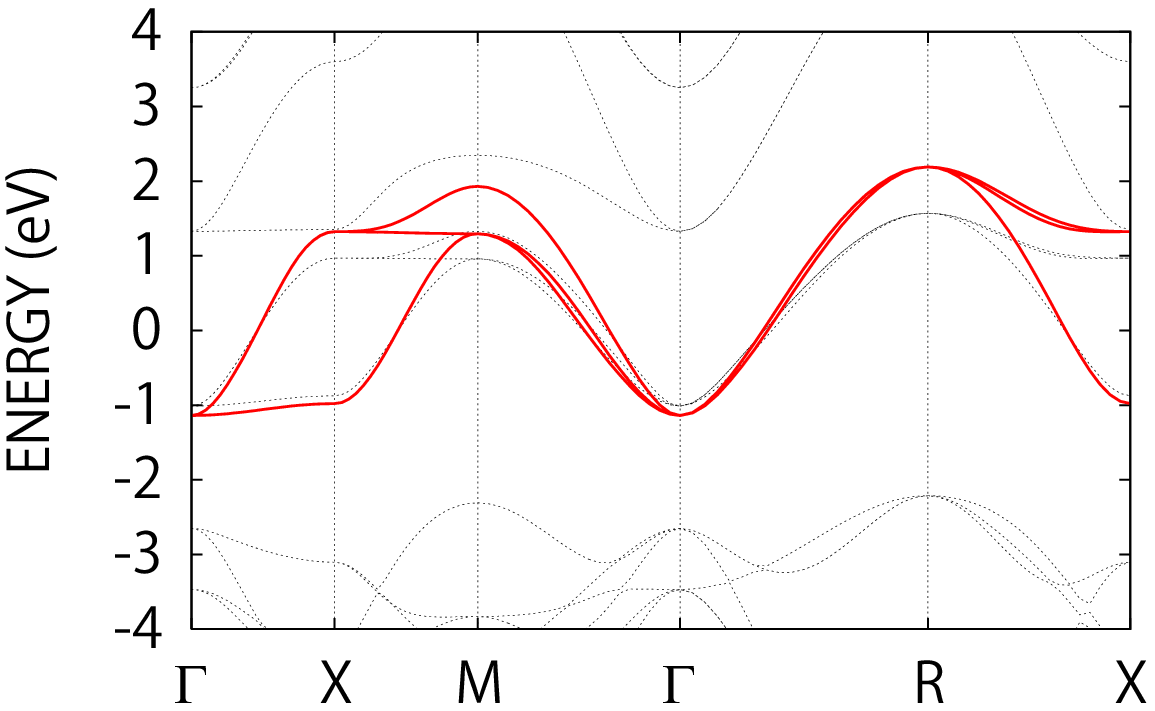} 
\caption{(Color online)
Band structure  of $[\epsilon_{\rm DFT}-V_{\rm xc}+\Delta\Sigma_{\rm H}+\Delta \Sigma_{\rm L}^{\rm nonlocal}]Z_{\rm HL}$ of SrVO$_3$ in the direct perturbative treatment.
For comparison, the LDA band structure is also given
(black dotted line).
The energy is measured from the Fermi level.
}
\label{DFT+Sigma_HL_Pert}
\end{figure} 
\begin{figure}[h]
\centering 
\includegraphics[clip,width=0.45\textwidth ]{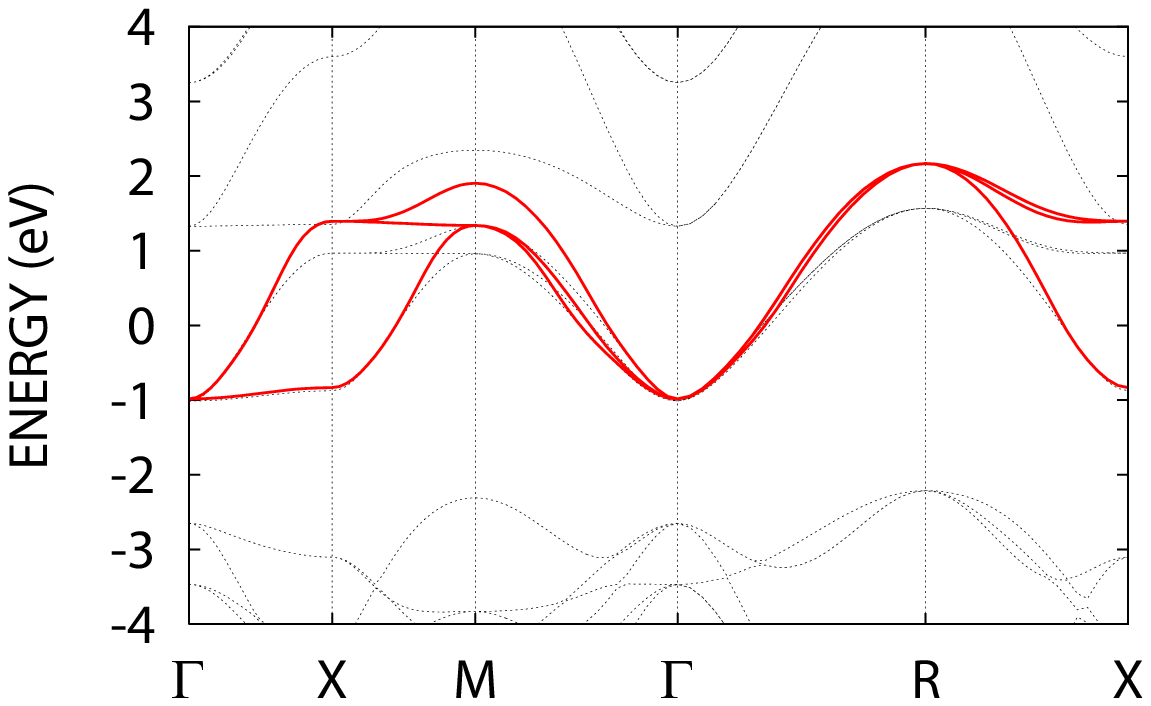} 
\caption{(Color online)
Band structure  of $[\epsilon_{\rm DFT}-V_{\rm xc}+\Delta\Sigma_{\rm H}+\Delta \Sigma_{\rm L}^{\rm nonlocal}]Z_{\rm HL}$ of SrVO$_3$ in the GW treatment.
For comparison, the LDA band structure is also given
(black dotted line).
The energy is measured from the Fermi level.
}
\label{DFT+Sigma_HL_GW}
\end{figure} 

The corresponding band structures 
are given by 
 $[\epsilon_{\rm DFT}+\Delta\Sigma_{\rm H}+\Delta \Sigma_{\rm L}^{\rm nonlocal}]Z_{\rm HL}$ in Fig.~\ref{DFT+Sigma_HL_Pert} 
for the perturbative treatment (see Eq.~(\ref{DeltaSigmaLnonlocalPert})), and in Fig.~\ref{DFT+Sigma_HL_GW} 
for the GW treatment,
where $\Delta \Sigma_{\rm L}^{\rm nonlocal}$ is defined in Eq.~(\ref{DeltaSigLnonlocalGW}) and $Z_{\rm HL}$ is around 0.92.
$Z_{\rm HL}$ is practically the same as $Z_{\rm H}$, meaning that the nonlocal dynamical correction 
in the H-space is small, which justifies the GW perturbative treatment for the nonlocal part.
One finds that 
the nonlocal part $\Delta \Sigma_{\rm L}=G^{(0)}_{ll}(W-W_{\rm L})$ narrows
the empty states but widens the occupied ones, resulting
in an overall bandwidth of 3.2 eV.
The difference between the direct perturbation and the GW treatment is small.

\begin{figure}[h]
\centering 
\includegraphics[clip,width=0.45\textwidth ]{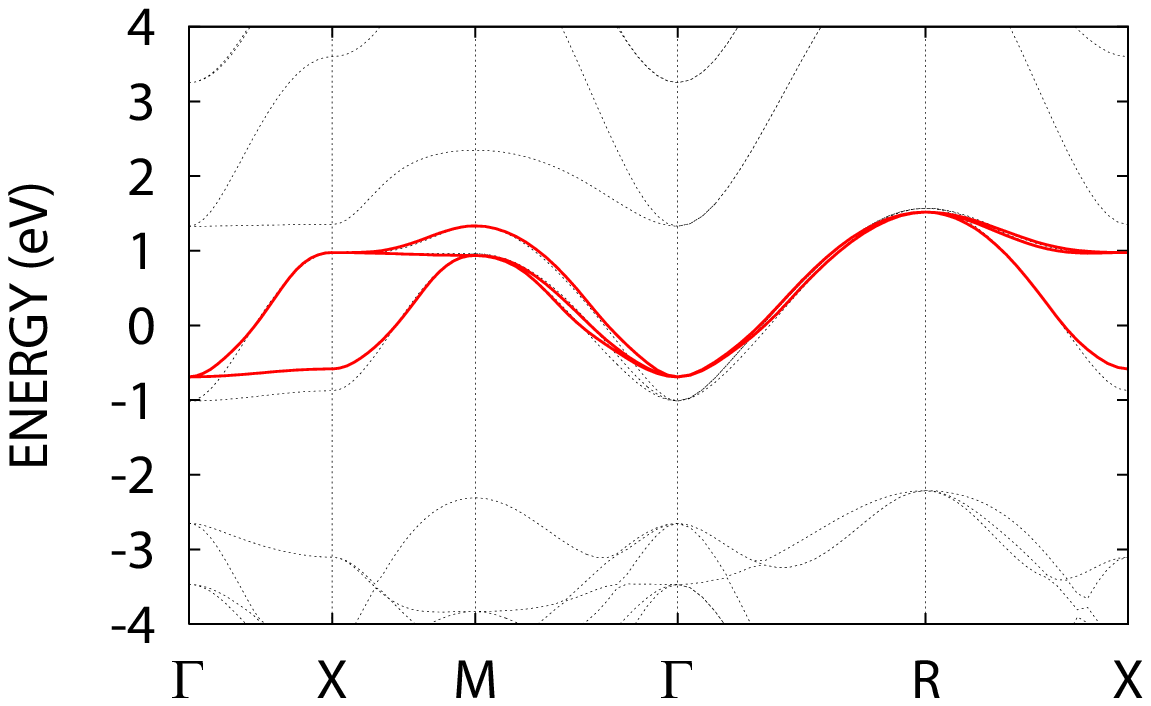} 
\caption{(Color online)
Band structure  of $[\epsilon_{\rm DFT}-V_{\rm xc}+\Delta\Sigma_{\rm H}+\Delta\Sigma_{\rm L}^{\rm nonlocal}]Z_{\rm HL}Z_B$ of SrVO$_3$ in the GW treatment.
For comparison, the LDA band structure is also given
(black dotted line).
The energy is measured from the Fermi level.}
\label{DFT+DeltaSigmaL_GW_ZHL_ZB}
\end{figure} 

After including the effect of the local self-energy by the nonperturbative 
Casula trick, we show the dispersion
given by 
$[\epsilon_{\rm DFT}-V_{\rm xc}+\Delta\Sigma_{\rm H}+\Delta\Sigma_{\rm L}^{\rm nonlocal}]Z_{\rm HL}Z_B$ for the GW-type treatment in Fig.~\ref{DFT+DeltaSigmaL_GW_ZHL_ZB}: 
The GW-like treatment gives a band dispersion
close to the LDA one. 
The $Z_B$ factor corresponding to the local dynamical part of 
$\Delta \Sigma_{\rm L}$ amounts to $Z_B=0.7$.
The nonperturbative treatment 
results in an LDA-like band dispersion for the empty states,
but a slightly narrower bandwidth in the occupied part.
The low-energy effective Hamiltonian at this level of treatment has a 
frequency independent effective interaction both with local 
and nonlocal interaction given by the Fourier transform of 
$W_{\rm H}(q,\omega=0)$, which contains both direct and exchange interactions. 

\begin{figure}[h]
\centering 
\includegraphics[clip,width=0.45\textwidth ]{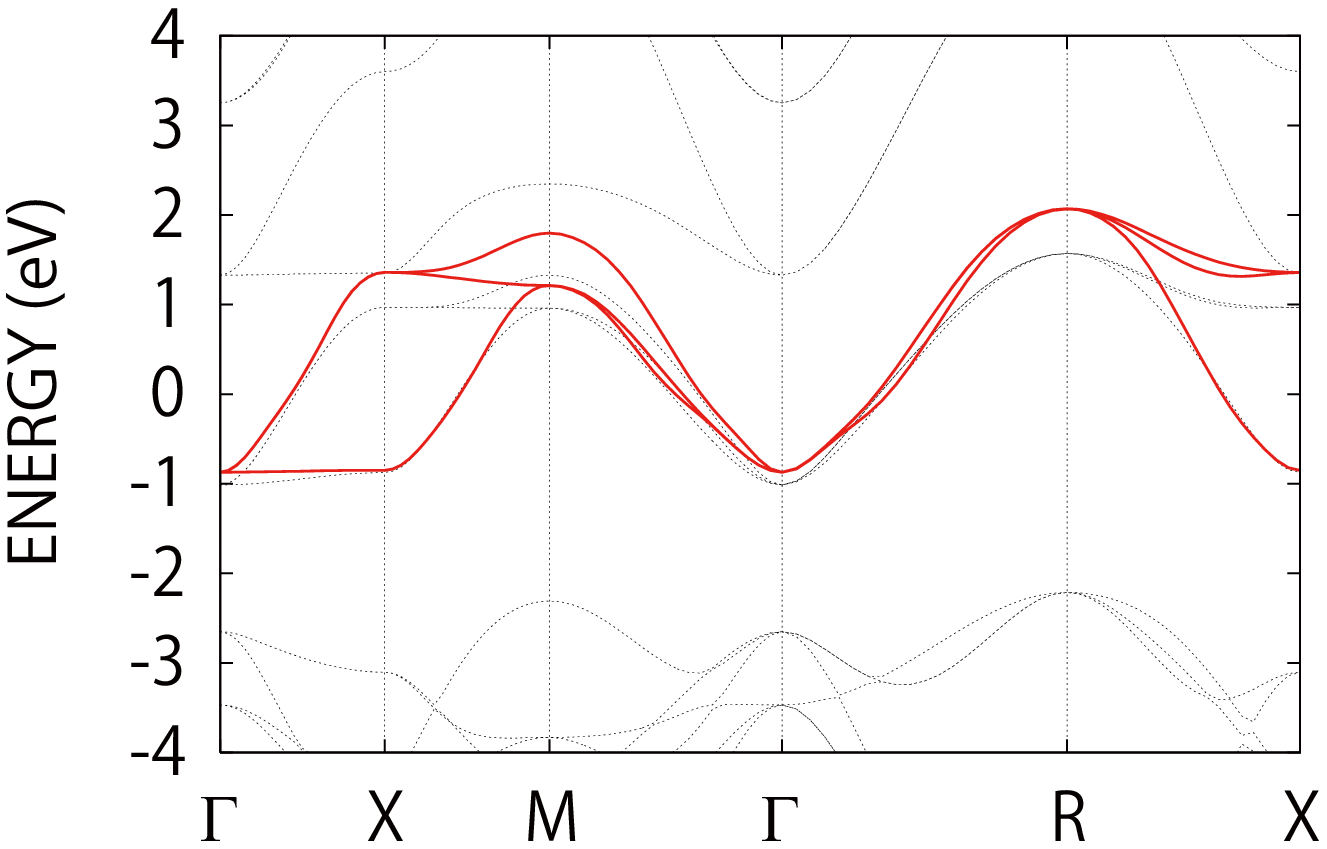} 
\caption{(Color online)
Band structure  of $[\epsilon_{\rm DFT}+\Sigma_{\rm corr}^{\rm GW}]Z_{\rm corr}Z_B$ of SrVO$_3$.
For comparison, the LDA band structure is also given
(black dotted line).
The energy is measured from the Fermi level.}
\label{DFT+Sigma_corr_GW_Z}
\end{figure} 

We finally show the dispersion
given by Eq.~(\ref{Teff_final}), namely
$[\epsilon_{\rm DFT}+\Sigma_{\rm corr}^{\rm GW}]Z_{\rm corr}Z_B$ in Fig.~\ref{DFT+Sigma_corr_GW_Z}. 
The result shows an LDA-like band dispersion for the occupied states,
but a slightly wider bandwidth in the empty part, resulting in a bandwidth 14\% wider in total than the LDA bandwidth. 
The effective interaction of the low-energy effective Hamiltonian 
at this final level has
frequency independent local and nonlocal interactions 
given by the Fourier transform of $W_r(q,\omega=0)$.

\begin{figure}[h]
\centering 
\includegraphics[clip,width=0.45\textwidth ]{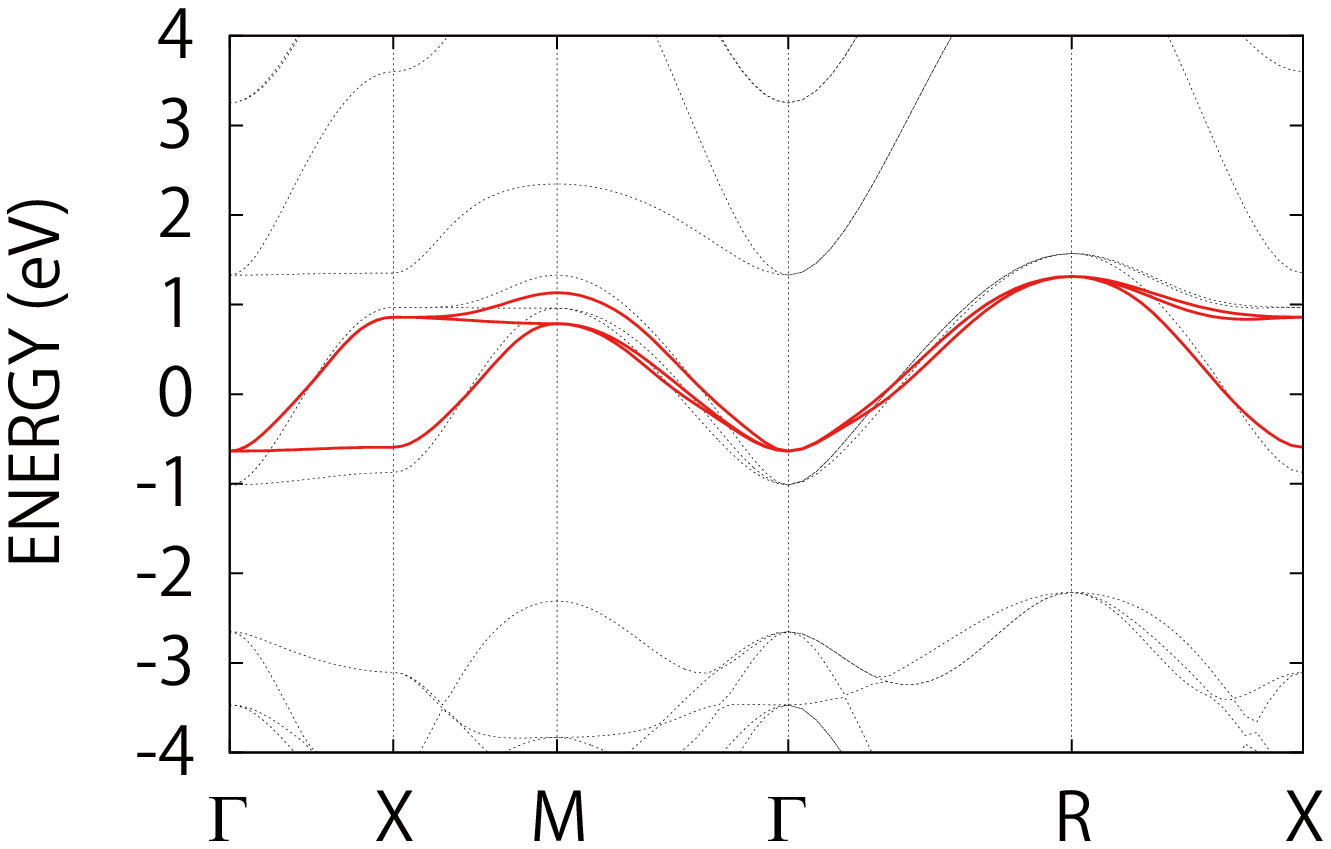} 
\caption{(Color online)
Band structure  of $[\epsilon_{\rm DFT}-V_{\rm xc}+\Sigma-(G^{(0)}W_{\rm GW}^{\rm dyn})^{\rm nonlocal}]Z'Z_B$ of SrVO$_3$
(Z' is the renormalization factor of $\Sigma-(G^{(0)}W_{\rm GW}^{\rm dyn})^{\rm nonlocal}$).
For comparison, the LDA band structure is also given
(black dotted line).
The energy is measured from the Fermi level.
}
\label{GW+ZB}
\end{figure} 

The resulting effective Hamiltonian in the L-space 
consists of single-particle and two-particle (interaction) parts.
The effective interaction is the same as in previous estimates 
by the cRPA in the literature, while the single-particle dispersion 
is revised after removing the double counting and taking into 
account the frequency dependence of the effective interaction. The 
final effective bandwidth for SrVO$_3$ is,
after partial cancellation, 
slightly (14\%) larger than the LDA bandwidth.  
This is a reasonable result because the LDA takes into account all 
correlations though insufficiently, while the present scheme by the constrained self-energy excludes the 
correlation effects arising from the L-space.
Although the bandwidth derived for the effective model is slightly 
larger than that of the LDA,  it is clear that the correlation effects 
are stronger than in the LDA or standard GW when the {\it ab initio} 
model is solved by an accurate solver. In fact, if the effective 
interaction contained in the final effective model is treated by 
the GW scheme, one obtains the dispersion illustrated in 
Fig.~\ref{GW+ZB}, 
which is given from the self energy of the whole GW calculation $\Sigma$ 
by correcting the local dynamical part $W-W_{\rm L}$ by $Z_B$. 
This indicates that even an insufficient treatment of the local 
static interaction by the GW scheme gives a dispersion with the bandwidth ($\sim 1.9$ eV) narrower 
than the LDA ($\sim 2.5$ eV, Fig.~\ref{lda}) and GW ($\sim 2.1$ eV, Fig.~\ref{fullGW}) dispersions.  
A slightly ($\sim 14$\%) wider dispersion than that of the LDA bands obtained for the 
effective Hamiltonian accompanied with frequency independent effective 
Coulomb interactions without the exchange part may account for the slight 
overestimate of correlation effects in the literature mentioned in 
the introduction.
Our scheme offers an optimized way for the  derivation of 
{\it ab initio} models for the L-space after eliminating the H-space
in  a systematic fashion.

Our findings of the band widening are consistent with studies
based on the combined GW+DMFT method in the literature 
\cite{TomczakPRB2014}: There, it was argued that within GW+DMFT
the best effective Hamiltonian that DMFT should be performed on,
is a one-body Hamiltonian corrected by the nonlocal part of the
GW self-energy. The corresponding spectral function 
(see e.g.
Figure 5 of Ref.~\onlinecite{TomczakPRB2014}) displays 
a broadening similar to 
Fig.~\ref{DFT+Sigma_HL_GW}.
Most interestingly, our present calculations confirm a pronounced
asymmetry observed in Ref.~\onlinecite{TomczakPRB2014}, namely a stronger widening
effect in the unoccupied part of the spectrum than in the occupied
one.

\section{Summary and Conclusion}
\label{conclusion}

In this work, we have developed and elaborated a method for a 
truly first principles electronic description of correlated
electron materials. Conceptually speaking, the scheme is based 
on renormalization group (RG) arguments, which guarantee the existence
of an effective Hamiltonian valid at a given energy scale.
The difficulty consists, however, in determining this Hamiltonian
explicitly, since a direct quantitative RG treatment of the full 
Coulomb Hamiltonian is a very difficult task and has not been
achieved so far. Indeed, performing RG calculations with long-range
interactions in the continuum is an even more difficult task
than for simplified models \cite{shankar}, and even for interacting 
lattice models explicit RG calculations remain a challenge.

Our scheme proposes an indirect way of constructing the 
low-energy Hamiltonian
which can be considered a shortcut to a true RG treatment.
The RG has to satisfy the chain rule, where the full trace summation denoted by ${\mathcal R}$ can be decomposed into the subsequent two partial trace ${\mathcal R}_{\rm H}{\mathcal R}_{\rm L}$ as required for the semigroup.
The first part ${\mathcal R}_{\rm H}$ can be replaced by a perturbative treatment in a controlled approximation because of the well separation of the L- and H- spaces.
Then the idea can be understood as working one's way backwards, starting
from the full solution obtained within some approximation (here, 
perturbation theory). The desired low-energy Hamiltonian is 
constructed such as to fulfill the following requirement:
its solution within the same approximation applied to the low-energy
subspace only should yield the same result as the projection of
the full solution to that subspace.
The motivation of this constructions resides in the fact that
instead of solving the resulting low-energy Hamiltonian within
the approximation used for its construction, more accurate
many-body solvers can be used for the final solution.

Strongly correlated electron systems provide a natural ground
for such a treatment, due to their hierarchical structure in
energy space, which facilitates the identification of appropriate
low-energy windows.
Nevertheless, in practice, the explicit construction of accurate 
low-energy effective Hamiltonians has remained a challenge due to the
difficulties associated to bridging the description of the
high-energy degrees of freedom usually treated in the DFT and 
the low-energy degrees of freedom described by the effective Hamiltonian
in a consistent manner. 
The main obstacles are related to
(1) the need of avoiding double counting of correlations 
and screening in the high- and low-energy treatments and (2) 
the frequency dependence of parameters in the low-energy effective models. 

In this work, we have presented a way to
overcome these bottlenecks: we propose a systematic recipe how
a low-energy Hamiltonian can be constructed by starting from a 
perturbative treatment.
We provide the best description under the constraint that the effective 
low-energy Hamiltonian contains only single-particle (kinetic energy term) and 
two-particle (interaction energy) terms with frequency independent parameters.  
Our construction relies on a controlled perturbative treatment,
which is possible thanks to the hierarchical nature of correlated
electron systems: even in cases, where perturbation theory would
not provide a meaningful approximation to the full problem, a
perturbative treatment of the high-energy degrees of freedom only
can be justified thanks to the fact that quantum many-body fluctuations
primarily live in the low-energy space only.

On the example of the ternary transition metal compound
SrVO$_3$, we have explicitly demonstrated how this
construction works: a low-energy Hamiltonian is built in such a way
that a perturbative treatment would 
reproduce the result of a perturbative treatment in the full
space as closely as possible. Solving the resulting many-body
Hamiltonian within accurate nonperturbative many-body solvers
then provides a description beyond the perturbative
treatment, while still keeping the {\it ab initio} nature of
the calculation.
We have tested our scheme in a step-by-step manner,
identifying the effects of the different corrective terms.
Most interestingly, our results confirm recent findings
within GW+DMFT calculations on SrVO$_3$ which identified an
intriguing asymmetry in the corrections to an LDA Hamiltonian
\cite{TomczakPRB2014}.
Our substantially improved MACE scheme should thus
open new ways to accurate many-body calculations beyond 
current {\it ab initio} methods.

\section{Acknowledgements}
We thank A. van Roekeghem for help with the xcrysden package.
This work was financially supported by MEXT
HPCI Strategic Programs for Innovative Research
(SPIRE) and RIKEN Advanced Institute for
Computational Science (AICS) through the HPCI System
Research Project (under Grants Nos. hp130007,
hp140215, and hp150211) and the Computational Materials Science
Initiative (CMSI).
This work was also supported by the European Research Council
under its Consolidator Grant scheme (project number 617196)
and IDRIS GENCI under project number t2015091393.

\section{Appendix}
In this Appendix, we analyse the difference between the local
GW self-energy
\begin{eqnarray}
\Sigma_{\rm loc} = G_{\rm loc} W_{\rm loc}
\end{eqnarray}
and the GW self-energy obtained from a local $W$.
\\
We use the expansions 
\begin{eqnarray}
G(r,r^{\prime}) = \sum_{R R^{\prime} L L^{\prime} }
G_{R R^{\prime} L L^{\prime} }
\chi_{RL} (r)
\chi_{R^{\prime}L^{\prime}} (r^{\prime})
\end{eqnarray}
and
\begin{eqnarray}
W(r,r^{\prime}) = \sum_{R R^{\prime} \alpha \beta }
W_{R R^{\prime}  \alpha \beta }
B_{R \alpha} (r)
B_{R^{\prime} \beta} (r^{\prime})
\end{eqnarray}
on the one- and two-particle bases $\chi$ and $B$,
respectively (following standard notations in the field).
Then, the $GW$ equation
\begin{eqnarray}
\Sigma(r, r^{\prime})  = G(r, r^{\prime}) W(r, r^{\prime}) 
\end{eqnarray}
leads to
\begin{eqnarray}
\Sigma_{R_1 R_2 L_1 L_2 }
= \sum_{R R^{\prime} L L^{\prime} }
\sum_{\tilde{R} \tilde{R}^{\prime} \alpha \beta}
G_{R R^{\prime} L L^{\prime} }
W_{\tilde{R} \tilde{R}^{\prime} \alpha \beta}
\nonumber
\\
\langle \chi_{R_1 L_1} \chi_{RL} | B_{\tilde{R^{\prime}} \alpha }\rangle
\langle B_{\tilde{R} \beta }|
\chi_{R_2 L_2}
\chi_{R^{\prime}L^{\prime}} 
\rangle
\end{eqnarray}
For a local $W$, that is an interaction of the form
$W_{\tilde{R} \tilde{R}^{\prime} \alpha \beta} \sim \delta_{RR^{\prime}}$
one has
\begin{eqnarray}
\Sigma_{R_1 R_2 L_1 L_2 } 
= \sum_{R R^{\prime} L L^{\prime} }
\sum_{\tilde{R} \alpha \beta}
G_{R R^{\prime} L L^{\prime} }
W_{\tilde{R} \tilde{R} \alpha \beta}
\nonumber
\\
\langle \chi_{R_1 L_1} \chi_{RL} | B_{\tilde{R} \alpha} \rangle
\langle B_{\tilde{R} \beta} |
\chi_{R_2 L_2}
\chi_{R^{\prime}L^{\prime}} 
\rangle
\end{eqnarray}
The structure of this equation is determined by the
overlap matrices of two-particle and one-particle basis
states. If the basis set is sufficiently localized that
no overlaps between basis functions on different spheres
need to be considered, these become local quantities
themselves: $O_{L_1 L \alpha} =
\langle \chi_{R L_1} \chi_{RL} | B_{R \alpha} \rangle$
and the above expression equals
the local self-energy
\begin{eqnarray}
\Sigma_{R R L_1 L_2 } 
= \sum_{ L L^{\prime} }
\sum_{ \alpha \beta}
G_{R R L L^{\prime} }
W_{R R \alpha \beta}
\nonumber
\\
\langle \chi_{R L_1} \chi_{RL} | B_{R \alpha} \rangle
\langle B_{R \beta} |
\chi_{R L_2}
\chi_{R L^{\prime}} 
\rangle
\end{eqnarray}
This is used in order to write Eq.~(\ref{DeltaSigmaLnonlocalPert_simple}).


\begin{thebibliography}{9}
\bibitem{footnote} As above, if partial self-consistency is prefered,
one may replace $G^{(0)}_{ll}$ by $G^{(2)}_{ll}$.
\end{thebibliography}
\end{document}